\documentclass[fleqn,usenatbib]{mnras}

\usepackage{newtxtext,newtxmath}
\usepackage{graphicx}% Include figure files
\usepackage{dcolumn}% Align table columns on decimal point

\usepackage[usenames,dvipsnames]{xcolor}
\hypersetup{colorlinks=true,citecolor=blue,linkcolor=OliveGreen}

\usepackage{soul}%highlighting {blah}

\usepackage{subfigure}
\usepackage{wasysym}
\usepackage{verbatim}
\usepackage{color}
\usepackage{mathtools}
\usepackage{hyperref}
\usepackage{soul}
\usepackage{pbox}
\usepackage[export]{adjustbox}

\usepackage{slashed}
\usepackage{multirow}
\usepackage{booktabs}  % for prettier-than-default tables
\usepackage{natbib}
\usepackage{float}
\usepackage{afterpage}
\usepackage{dcolumn}
\usepackage{bm}
\usepackage{lipsum}
\usepackage{setspace}
\usepackage{etoolbox}

\def\beq{\begin{equation}}
\def\eeq{\end{equation}}

\def\mnras{Mon. Not. R. Astron. Soc}
\def\physrep{Physics Reports}

\usepackage[T1]{fontenc}
\usepackage{ae,aecompl}

\usepackage{graphicx}	% Including figure files
\usepackage{amsmath}	% Advanced maths commands
\title[\textsc{Galaxy distribution in IllustrisTNG and a semi-analytic model}]{Galaxy assembly bias and large-scale distribution: a comparison between IllustrisTNG and a semi-analytic model}

\author[B. Hadzhiyska et al.]{
Boryana Hadzhiyska,$^{1}$\thanks{E-mail: boryana.hadzhiyska@cfa.harvard.edu}
Sonya Liu,$^{2}$
Rachel S. Somerville,$^{3}$
Austen Gabrielpillai,$^{3}$
Sownak Bose,$^{1}$
\newauthor 
\ Daniel Eisenstein,$^{1}$
Lars Hernquist,$^{1}$
\\
$^{1}$Center for Astrophysics $\vert$ Harvard \& Smithsonian, 60 Garden St., Cambridge, MA 02138, USA\\
$^{2}$American Heritage School, 736 N 1100 E, American Fork, UT 84003, United States\\
$^{3}$Center for Computational Astrophysics, Flatiron Institute, 162 5th Avenue, New York, NY 10010
}
\date{Accepted XXX. Received YYY; in original form ZZZ}

\pubyear{2021}

\begin{document}
\label{firstpage}
\pagerange{\pageref{firstpage}--\pageref{lastpage}}
\maketitle

% Abstract of the paper
\begin{abstract}
In this work, we compare large scale structure observables for stellar mass selected samples at $z=0$, as predicted by two galaxy models, the hydrodynamical simulation IllustrisTNG and the Santa-Cruz semi-analytic model (SC-SAM). Although both models have been independently calibrated to match observations, rather than each other, we find good agreement between the two models for two-point clustering and galaxy assembly bias signatures. The models also show a qualitatively similar response of occupancy and clustering to secondary halo paramaters other than mass, such as formation history and concentration, although with some quantitative differences. Thus, our results demonstrate that the galaxy-halo relationships in SC-SAM and TNG are quite similar to first order. However, we also find areas in which the models differ. For example, we note a strong correlation between halo gas content and environment in TNG, which is lacking in the SC-SAM, as well as differences in the occupancy predictions for low-mass haloes. Moreover, we show that higher-order statistics, such as cumulants of the density field, help to accurately describe the galaxy distribution and discriminate between models that show degenerate behavior for two-point statistics. Our results suggest that SAMs are a promising cost-effective and intuitive method for generating mock catalogues for next generation cosmological surveys.

\end{abstract}

% Select between one and six entries from the list of approved keywords.
% Don't make up new ones.
\begin{keywords}
cosmology: large-scale structure of Universe -- galaxies: haloes -- methods: numerical -- cosmology: theory
\end{keywords}

%%%%%%%%%%%%%%%%%%%%%%%%%%%%%%%%%%%%%%%%%%%%%%%%%%

%%%%%%%%%%%%%%%%% BODY OF PAPER %%%%%%%%%%%%%%%%%%

\section{Introduction}
\label{sec:intro}
According to the concordance model of Big Bang cosmology, our Universe is made up of a network of filaments, shaped by gravity and populated with dark matter haloes. These dark matter haloes correspond to overdense regions, which have evolved by gravitational instability and interactions with other haloes. In this framework, galaxy formation takes place within such haloes, as baryonic matter sinks to the centre of their potential wells and cold gas condenses to form galaxies \citep{1978MNRAS.183..341W}. Since we cannot observe dark matter directly, understanding the galaxy-halo connection by studying the distribution of galaxies would enable us to place stringent constraints on cosmological parameters.

The formation and evolution of dark matter haloes can be modeled with a high degree of accuracy by cosmological ($N$-body) simulations. Such calculations incorporate various assumptions of the standard paradigm and adopt particular values for the cosmological parameters, which are often determined by observations. They have the benefit of being relatively computationally inexpensive and, thus, can encompass very large volumes ($\sim$1 Gpc$/h$); a necessity to perform high-fidelity forecasts and interpretation of current and future galaxy surveys. However, a considerable shortcoming of these simulations is that they lack prescriptions for determining the detailed formation and evolutionary history of the observable components of galaxies. To mimic the galactic component in such large volumes, cosmologists often adopt \textit{a posteriori} models of varying complexity to populate the dark matter haloes with galaxies. 

The simplest of these models involve phenomenological approaches, such as halo occupation distribution (HOD) models, which provide an empirical relation between halo mass and the number of galaxies in a host halo \citep{2000MNRAS.311..793B,2000MNRAS.318.1144P,2001ApJ...546...20S,2004MNRAS.350.1153Y,2018ARA&A..56..435W}. HOD modelling is one of the most popular ways to  ``paint'' galaxies onto $N$-body simulations. The process of generating multiple mock catalogues in large volumes is very efficient and can thus satisfy the demanding requirements of next-generation experiments. Unfortunately, the simplest versions of such empirical approaches have been shown to make predictions that have significant discrepancies with observations and more detailed galaxy formation models \citep{2007MNRAS.374.1303C,2015MNRAS.454.3030P,2020MNRAS.491.5771B}. For example, several recent works have argued that including secondary HOD parameters, such as the local halo environment and concentration, yields better agreement with hydrodynamical simulations as compared to the standard prescription \citep{2020MNRAS.493.5506H,2020MNRAS.492.2739X,2021MNRAS.501.1603H}. 
Related approaches include Sub-Halo Abundance Matching models (SHAMs) and semi-empirical models, which derive mappings between galaxy and DM halo properties such that specific observational quantities are matched \citep{2013ApJ...770...57B,2013ApJ...768L..37T, 2013MNRAS.428.3121M,2018MNRAS.477.1822M,2018ApJ...868...92T,2019MNRAS.488.3143B}. These methods can also be effective for efficient construction of realistic galaxy mock catalogues. A common feature of HOD, SHAM, and semi-empirical models is that they derive empirical mappings between galaxies and DM halos, without attempting to model \textit{ab initio} physical processes. 

On the other hand, numerical hydrodynamical simulations provide an \textit{ab initio} approach for gaining insight into the evolution and distribution of galaxies \citep[e.g.][]{2014MNRAS.444.1518V,2015MNRAS.446..521S}, incorporating a broad suite of physical processes, including gravity, hydrodynamics, and thermodynamics. These simulations also incorporate key baryonic processes such as star formation, stellar winds, chemical enrichment, and black hole growth and feedback. However, due to the limited dynamic range of the simulations, such processes must be treated with "sub-grid" recipes which describe physics that occurs below the explicit resolution of the numerical scheme. These sub-grid processes are typically phenomenological, and contain adjustable parameters which are typically calibrated to reproduce a set of observations. Having recently reached sizes of several hundred megaparsecs on a side \citep{2016MNRAS.460.3100C,2018MNRAS.475..676S}, hydro simulations now allow us to test the accuracy of galaxy population models by comparing various statistical properties of the galaxy field side by side \citep{sownak,2020MNRAS.493.5506H,2021MNRAS.504.5205C}. Such analyses have been performed extensively for galaxy populations above a certain stellar mass cut (as stellar-mass cuts are a good proxy for luminosity cuts), and a few  studies have focused on other galaxy populations targeted by future surveys, such as star-forming emission-line galaxies (ELGs) \citep[e.g.][]{Zheng:2004id,2018MNRAS.480..864C,2017MNRAS.472..550F,2016MNRAS.459.3040G,2018MNRAS.474.4024G,2019MNRAS.483.4501A,2021MNRAS.502.3599H}. 
%rss seems off topic for this paragraph
%Developing more robust models of the galaxy-halo connection for star-forming and bright luminous galaxies is especially important as they will constrain cosmological parameters in future experiments. 
While hydro simulations can be used to inform us about the galaxy-halo connection on intermediate and quasi-linear scales, they are computationally expensive and cannot (at present) be run in the large volumes needed for future cosmological surveys. Moreover, the predictions for galaxy properties are sensitive to the details of the sub-grid prescriptions, and it is too expensive to explore many variations of these in large volumes. 

%rss to me, these "semi-empirical" models are really more like SHAMs than SAMs. so moving these references up with the HOD paragraph
%As an intermediate approach, there are more complex empirical models that study the relationship between galaxy properties and their host haloes, which can potentially aid the construction of realistic galaxy mock catalogues \citep{2013ApJ...770...57B,2013ApJ...768L..37T, 2013MNRAS.428.3121M,2018MNRAS.477.1822M,2018ApJ...868...92T,2019MNRAS.488.3143B}. 
A particularly promising and well-explored route for physics-based yet more computationally efficient modeling of the galaxy-halo connection is through semi-analytic models (SAMs) of the galaxy formation. SAMs are built using halo merger trees extracted from an $N$-body simulation and employ various physical prescriptions to predict the evolution of baryons and their flow between different ``reservoirs", such as the intergalactic medium (IGM), circumgalactic medium (CGM), interstellar medium (ISM), and galactic stellar bulge, disk, and halo \citep[for a review, see][]{2015ARA&A..53...51S}.  A caveat of these models is that the evolution of baryonic matter is not obtained by directly solving the equations of gravity, hydrodynamics, thermodynamics, etc, but rather approximated through analytic calculations. Similarly to cosmological hydro simulations, key physical processes such as star formation, stellar feedback, and BH feedback are treated with parameterized phenomenological recipes, which are calibrated to match a subset of observations.

Understanding the connection between galaxies and their host dark matter haloes will help us obtain more realistic mock catalogues, which are crucial for the analysis of future observational surveys. One of the main limitations of ubiquitously adopted galaxy-halo approaches, such as the HOD model, is the amount of ``galaxy assembly bias'' they predict. Galaxy assembly bias is the relative increase (or decrease) of the large-scale clustering of a galaxy sample resulting from dependencies of the galaxy-halo connection on halo properties beyond mass, such as assembly history and environment \citep[e.g.][]{2007MNRAS.374.1303C}. Thus, the galaxy assembly bias signal is the consequence of the effects of both halo assembly bias and occupation variations. The former is the dependence of halo clustering on properties other than mass \citep[e.g.][]{2005MNRAS.363L..66G}. The latter is the dependence of the properties of the galaxies within a halo on halo properties in addition to mass \citep[see][]{2018ApJ...853...84Z,2018MNRAS.480.3978A}. Some of the properties thought to contribute to the galaxy assembly bias signal are formation time, concentration, spin, and environment. Various authors have compared observations and realistic galaxy models with mock catalogues generated with and without taking assembly bias into consideration in an effort to study the strength of that effect \citep{2019MNRAS.484.1133C}.

In this paper, we explore the large-scale galaxy distributions predicted by the numerical hydrodynamic simulation IllustrisTNG \citep{2017MNRAS.465.3291W,2018MNRAS.473.4077P} and the Santa-Cruz semi-analytic model \citep{somerville:1999,2008MNRAS.391..481S,2015MNRAS.453.4337S,somerville:2021,Austen+2021}.
%\citep[SC-SAM, see][for more details on the SC-SAM]{Austen+2021}. We investigate the manifestation of the assembly bias signal in both models and in particular, focus on understanding the effects of halo parameters other than mass on the clustering, thus gaining insights into the possible physical origin of galaxy assembly bias. Explaining the statistical properties of the large-scale galaxy distribution of TNG through a semi-analytic approach could greatly improve our ability to create realistic mock catalogues for the large volumes demanded by current and future experiments, such as SDSS, DESI, and \textit{Euclid} \citep{2016ApJS..224...34P,2016arXiv161100036D,2011arXiv1110.3193L}. 

This paper is structured as follows. In Section \ref{sec:meth}, we summarize the semi-analytic model adopted in this work, SC-SAM, and briefly describe the IllustrisTNG suite of simulations. Moreover, we define various secondary halo parameters considered in our study of the assembly bias signal as well as the galaxy samples for which we compute summary statistics. In Section \ref{sec:res}, we compare the large-scale properties of the TNG and SAM samples. In particular, we explore the halo occupation distributions, the clustering and the assembly bias of our galaxy samples, and the variation caused by secondary parameters. We also examine the spatial distribution of the galaxies and the amount of gas in the host halo conditioning on its environment. Finally, we analyse the differences in the clustering properties of galaxies when using alternative statistical probes, such as galaxy-galaxy lensing, galaxy-void correlation functions and the cumulants of the smoothed galaxy density. We discuss the interpretation and implications of our results in Section~\ref{sec:discussion}, and our conclusions are summarized in Section \ref{sec:conc}.

\section{Methodology}
\label{sec:meth}
In this section, we describe the Santa-Cruz semi-analytic model and the hydrodynamical simulation, IllustrisTNG, used in this work. The main focus of this study is on mass-selected galaxy samples at $z = 0$, as predicted by both models. Other types of galaxy tracers at early redshifts will be explored in future work.

\subsection{Santa Cruz semi-analytic model}
\label{sec:sam}

The version of the Santa Cruz SAM adopted in this study is the same as \citet{Austen+2021} and is similar to SAMs published in \citet{2015MNRAS.453.4337S} and \citet{somerville:2021}. Interested readers are pointed to these references, as well as \citet{2008MNRAS.391..481S} for more information.

The semi-analytic model is run on merger trees created using the \textsc{ROCKSTAR} halo-finding algorithm \citep{2013ApJ...762..109B}. \textsc{ROCKSTAR} is a temporal, phase-space finder; due to its use of both the phase space distribution of particles and temporal evolution, it is considered highly accurate in determining particle-halo membership. Having information about the relative motion of haloes makes the process of finding tidal remnants and determining halo boundaries substantially more effective, while having temporal information helps maximize the consistency of halo properties across time, rather than just within a single snapshot. The parameters chosen to construct the merger trees are the default \textsc{ROCKSTAR} parameters \footnote{For more information, see \url{https://github.com/yt-project/rockstar/blob/master/README.md}.}. The \textsc{ROCKSTAR} merger trees provide information on the growth of collapsed dark matter haloes over time through accretion of ``diffuse'' material and mergers with other haloes.

The rate of gas flow into a halo is given by the product of the universal baryon fraction and the growth rate of the dark matter halo, i.e. $f_b \dot M_h$. This relation holds in the absence of feedback processes and takes into account the suppression of gas inflow due to photoionization through the expression for the collapsed baryon fraction:
\begin{equation}
  f_{\rm b, coll} = \frac{f_b}{[1+0.26 M_F(z)/M_{\rm vir}]^3},    
\end{equation} 
where $M_{\rm vir}$ is the halo virial mass and $M_F$ is the filtering mass, which depends on the reionization history of the Universe. The accreted gas forms the so-called ``hot halo" and is assumed to be isothermal and at the virial temperature. The ``cooling radius'' is computed using standard radiative cooling functions from \citet{1993ApJS...88..253S}. When the cooling radius is smaller than the virial radius, $r_{\rm vir}$, a standard cooling flow model is applied. The cooling time is calculated via  Eqn. 1 in \citet{2008MNRAS.391..481S}
\begin{equation}
   t_{\rm cool} = \frac{(3/2) \ \mu m_{\rm p} \ k T}{\rho_g(r) \ \Lambda(T, Z_{\rm h})} ,
\end{equation}
where $\mu m_{\rm p}$ is the mean molecular mass, $k T \propto V_{\rm vir}^2$ is the virial temperature, $\rho_{\rm g} = m_{\rm hot}/(4 \pi r_{\rm vir} r^2)$ is the gas density profile, and $\Lambda (T, Z_{\rm h})$ is the temperature and metallicity-dependent cooling function. As the cooling radius moves outward, the gas contained within it is assumed to cool and accrete onto the disk. 
When the cooling radius is larger than $r_{\rm vir}$, the cooling rate is given by the gas accretion rate into the halo. 

The star formation rate of galaxies is computed by assuming a scaling relation between the molecular gas surface density and star formation rate surface density, as supported by observations. The gas profile is assumed to be an exponential disc, proportional to the stellar disc, $r_{\rm disc}$, where $r_{\rm disc}$ is computed by assuming angular momentum conservation and utilising the halo parameters for concentration $c_{\rm NFW}$, spin $\lambda$, and the fraction of baryons in the disc $f_{\rm disc}$. Stellar feedback can eject cold gas from the ISM, with a rate determined by the circular velocity of the halo at twice the Navarro-Frenk-White \citep[NFW,][]{1996ApJ...462..563N} scale radius, $V_{\rm disc}$, the star formation rate $\dot m_{\ast}$, and adjustable parameters ($\epsilon_{\rm SN} = 1.7$, $\alpha_{\rm rh} = 3$): 
\begin{equation}
  \dot m_{\rm eject} = \epsilon_{\rm SN} \left( \frac{200 \ {\rm km/s}}{V_{\rm disc}} \right)^{\alpha_{\rm rh}} \dot m_{\ast} .    
\end{equation}
A fraction of this gas is ejected from the halo and stored in an ``ejected" reservoir, while the rest is deposited in the hot gas halo. Gas in the ejected reservoir is
``re-accreted'' at rate dependent on the halo dynamical time, $t_{\rm dyn} = r_{\rm vir}/V_{\rm vir}$, with a normalization that contains another adjustable parameter.

Each top-level halo is seeded with a black hole, which can accrete mass through cooling flows from the hot halo or by inflows of cold gas driven by mergers or internal gravitational instabilities. Cold gas can be ejected from the ISM through AGN-driven winds triggered by mergers or disk instabilities.  The radiately inefficient mode of growth (``radio mode'') generates a heating term, which can partially or completely offset cooling and accretion from the hot halo.
%at a rate given by \begin{equation}
%    \dot m_{\rm radio} \propto \frac{kT}{\Lambda(T, Z_{\rm h})} M_{\rm BH},
%\end{equation} 
%with the black hole mass $M_{\rm BH}$ proportional to the spheroid stellar mass. 
For more details on the physics recipes adopted in SC-SAM, see \citet{2008MNRAS.391..481S}.

Galaxies residing at the centre of their host haloes are considered ``centrals'', while those hosted by subhaloes, as defined by \textsc{ROCKSTAR}, are called ``satellites''. Rather than relying on N-body simulations for information about the subhaloes, the Santa-Cruz SAM tracks the orbital decay and tidal destruction of satellite galaxies utilising a semi-analytic prescription \citep[see][for details]{2008MNRAS.391..481S}.

The SC SAM is tuned to match a sub-set of observations at $z=0$ only, which are: the stellar mass function (equivalent to matching the stellar mass vs. halo mass relation), the stellar mass vs. cold gas fraction, stellar mass vs. metallicity relation, and the bulge mass vs. black hole mass relation. Please see \citet{Yung:2019a} Appendix~B and \citet{Austen+2021} for details of the observations used for calibration. 

An additional product of the SC-SAM runs presented in \citet{Austen+2021} is a catalogue of halo and subhalo bijective matches between IllustrisTNG and SC-SAM, which plays a crucial role in determining the dependence of large-scale clustering observables on the halo assembly bias of both models. We also utilise these matches to provide more realistic positions to the SC-SAM satellites, since as noted above, the SAM does not use the sub-halo positions and velocities from the N-body simulation. By taking the subhalo positions for a given \textsc{ROCKSTAR} halo directly from its bijectively matched FoF group, the positions of satellites in the SC-SAM correspond to the true positions of subhaloes belonging to that halo, as identified in TNG. The effect of this exercise is two-fold: on one hand, it simplifies the comparison of the one-halo term in TNG to SC-SAM; on the other, it enables us to examine the weak lensing signal of the two models, as the SC-SAM satellite distribution is correlated with the dark-matter distribution.

\subsection{IllustrisTNG}
\label{sec:tng}
The Next Generation Illustris simulation (IllustrisTNG, TNG), which is run with the AREPO code \citep{2010MNRAS.401..791S,2019arXiv190904667W}, consists of 9 simulations: 3 box sizes (300, 100 and 50 Mpc on a side), each available at 3 different resolutions, 1--3, with 1 being the highest and 3 the lowest resolution \citep{2018MNRAS.475..676S,2018MNRAS.477.1206N,2018MNRAS.480.5113M,2019MNRAS.tmp.2010N,2019MNRAS.tmp.2024P,2019MNRAS.490.3196P,2019MNRAS.490.3234N}. Compared with its predecessor, Illustris \citep{2014MNRAS.444.1518V,2014Natur.509..177V,2014MNRAS.445..175G},  TNG provides improved agreement with observations by modifying its treatment of active galactic nuclei (AGN) feedback,  galactic winds and magnetic fields \citep{2018MNRAS.473.4077P,2017MNRAS.465.3291W}. In addition, various improvements of the flexibility and hydrodynamical convergence have been introduced in the code.

%rss add some more details about physical processes in TNG?

The parameters that characterize the sub-grid processes in TNG have been calibrated to approximately reproduce a set of observations. Although the philosophy is similar to that used to calibrate SAMs, the details of which observations are used for calibration, and the required precision of the calibration, can be different. The observations used for calibration of TNG are discussed in \citet{2018MNRAS.473.4077P}, and include the cosmic SFR density as a function of redshift, stellar mass function, BH mass vs. stellar mass relation, hot gas fraction in galaxy clusters, and galaxy stellar mass vs. radius relation. In addition, the distribution of galaxy optical colors in stellar mass bins was also used to motivate and calibrate the AGN feedback parameters as discussed in \citet{2018MNRAS.475..624N}. We note that these choices overlap with, but differ from, those used to calibrate the SC SAM.

In this paper, we use the largest box ($L_{\rm box} = 205 \ \rm{Mpc}/h$) at its highest resolution, TNG300-1, with particle mass of $5.9 \times 10^7 M_{\odot}$ and $1.1 \times 10^7 M_{\odot}$ for the dark matter and baryons, respectively. TNG provides both the hydrodynamical (full-physics) and the dark matter only ($N$-body) simulation output, evolved from the same set of initial conditions. It is crucial to test any hypotheses regarding the galaxy-halo connection on the dark-matter-only counterpart, since mock catalogues used in cosmological galaxy surveys typically utilise cosmological $N$-body simulations, as they can be produced in sufficiently large volumes. Having a bijective match between the full-physics and $N$-body runs also gives us an opportunity to make halo-by-halo comparisons and thus study the effect of baryons on the properties of haloes.

Haloes (groups) in TNG are identified using the standard friends-of-friends \citep[FoF,][]{1985ApJ...292..371D,1987ApJ...319..575B} algorithm, which depends on a single parameter, the linking length $b = 0.2$, and is run on the dark matter particles. The outer boundaries of the FoF haloes roughly correspond to an overdensity contour of $\sim$180 times the background density using the percolation theory results of \citet{2011ApJS..195....4M}. However, haloes identified by this method may appear as two or more clumps, linked by a small thread of particles, which leads to a miscalculation of their properties. The subhaloes, on the other hand, are found using the SUBFIND algorithm \citep{Springel:2000qu}, which detects substructure within the groups and defines locally overdense, self-bound particle groups.

\subsection{Selection of the galaxy samples}
\label{sec:sample}
In this work, we consider stellar-mass-selected galaxy samples at $z = 0$. Although such mass-selected samples do not necessarily satisfy the same set of magnitude and color cuts that luminous red galaxies (LRGs) must satisfy to be targeted by modern galaxy surveys (e.g. DESI, SDSS, Euclid), they are believed to largely overlap with the LRGs selected in imaging surveys \citep{2013MNRAS.433.3506M}. 
%%rss added more info about the calibration to the SAM and TNG sections
%The SC-SAM employed in this study has been calibrated to match observations at $z = 0$. 
%A more complex tuning of the SAM that accounts for the redshift dependence of the matched properties and defines multiple tracer types (e.g. emission-line galaxies or ELGs) is deferred to future work.
%
The stellar-mass selection is performed at three different galaxy number densities, aimed to mimic current and near-future galaxy surveys, $n_{\rm gal} = 0.0005, \ 0.001, \ {\rm and} \ 0.002 \ [{\rm Mpc}/h]^{-3}$. The TNG subhalo stellar mass definition found to most closely correspond to the stellar mass reported in SC-SAM is `SubhaloMassInRadType'; i.e. the sum of masses of all `star' particles within twice the stellar half mass radius. The lowest stellar mass of the TNG sample is $\log M_\ast = 10.56$, while that of SC-SAM is $\log M_\ast = 10.81$, expressed in units of $M_\odot/h$. The TNG halo mass that we adopt is `Group\_M\_TopHat200' ($\equiv M_{\rm 200}$), i.e. the total mass enclosed in a sphere whose mean density is $\Delta_c$ times the critical density of the Universe. $\Delta_c$ is derived from the solution of the collapse of a spherical top-hat perturbation. A well-known fitting function is provided by
\begin{equation}
  \Delta_c(z) = 18 \pi^2 + 82 x - 39 x^2,
\end{equation} 
where $x = \Omega_m(z) - 1$ and $\Omega_m(z)$ is the matter energy density at redshift $z$ \citep{1998ApJ...495...80B}. The virial mass of \textsc{ROCKSTAR} haloes is defined using the same redshift scaling of $\Delta_c$.

We construct the TNG galaxy sample from the most massive $N$ galaxies, where $N = 6000, \ 12000, \ {\rm and} \ 24000$, corresponding to the three density thresholds above. This results in roughly a 3:1 split of centrals to satellites as defined by SUBFIND. Rather than utilising the same indiscriminate selection in SC-SAM, we require that the satellite-central ratio be the same in SC-SAM when selecting the sample of $N$ galaxies, ordered by stellar mass. Thus, we obtain two galaxy samples that have matching satellite fractions and galaxy densities. The constraint of a fixed satellite fraction is enforced due to a fundamental difference between the two models in the treatment of satellites. First, the merging, tidal stripping, and destruction of satellites in the SC-SAM is treated using a semi-analytic model. Second, once a halo becomes a sub-halo (satellite) in the SAM, it no longer experiences any new cooling from the CGM, since the satellite hot halos are assumed to be stripped instantaneously and added to the CGM of the central/host halo. Thus, satellites in SAMs may tend to be "over-quenched", i.e., they may cease forming stars earlier than satellites of the same mass and host mass in hydrodynamic simulations. 
%While satellite galaxies in TNG may continue to grow in mass throughout their history, satellites in SC-SAM are quenched the moment they are recognised as such. Since this affects the star-formation-rate (SFR) of satellites in the two models, we refrain from discussing SFR in this work. 
An additional secondary effect comes from the difference in the halo finders used to define haloes in both samples -- TNG uses FoF, while SC-SAM uses \textsc{ROCKSTAR} (see Section~\ref{sec:sam}). Haloes identified as two separate haloes with \textsc{ROCKSTAR} can often considered to be a single FoF halo, which affects the halo mass function, the response to halo assembly bias, and the classification of dark-matter structure into haloes and subhaloes. These effects are considered in App.~\ref{app:halos}, where we demonstrate that they do not influence the main conclusions in this work.

\begin{figure}
\centering  
\includegraphics[width=0.48\textwidth]{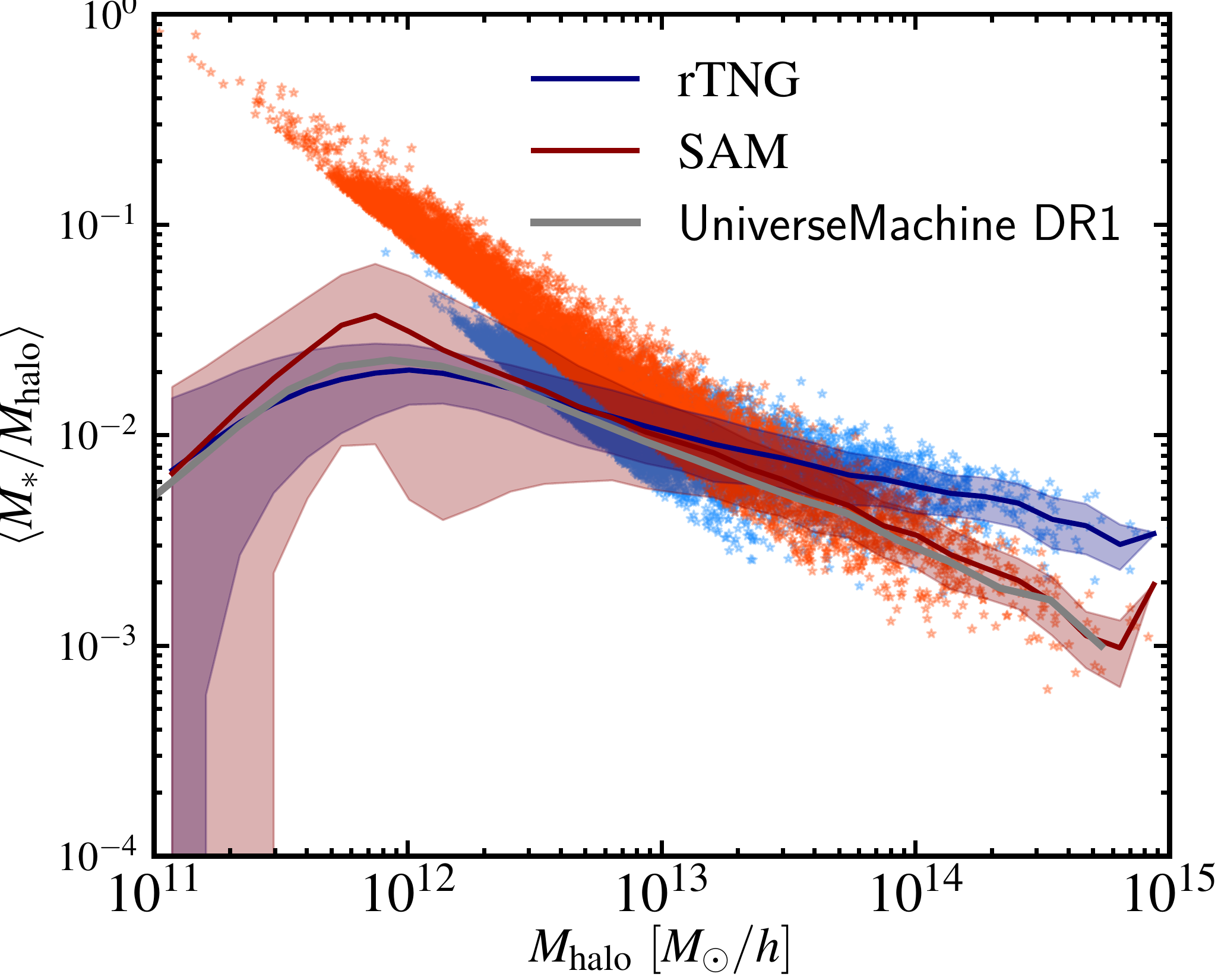}
\caption{Stellar-to-halo mass ratio for all galaxies in TNG and SC-SAM (blue and red lines and shaded areas) as well as for the two stellar mass selected samples defined in the text (blue and red dots; see Section~\ref{sec:sample}). The solid lines show the mean of the relation, while the shaded regions mark one standard deviation above and below the mean.
%\rss{what do the lines and shaded regions show exactly? %mean/median and some percentiles (which)?}
In solid gray, we show the UniverseMachine (UM) DR1 prediction and find that while the rescaled TNG (rTNG) galaxies 
%\citep{2018MNRAS.475..648P} 
provide a very good match for smaller haloes, $M_{\rm halo} \lesssim 10^{13} \ M_\odot/h$, on the high mass end, the TNG centrals appear to overproduce stars. On the other hand, SAM galaxies living in haloes with masses $M_{\rm halo} \gtrsim 10^{13} \ M_\odot/h$ exhibit a better agreement with UM, but show a discrepancy around the peak, $M_{\rm halo} \sim 10^{12} \ M_\odot/h$. 
%This discrepancy is also reflected in the selected sample.
%The SC SAM also predicts a much larger scatter in this relation around the peak, which results in the selection of 
%\rss{i don't think it makes sense to show the dashed %lines. also, the plot is a bit hard to read because the %red line blends in with the red points.}
}
\label{fig:shmr_scatter}
\end{figure}

In Fig. \ref{fig:shmr_scatter}, we show the stellar-to-halo mass ratio of central galaxies in all haloes for both models. 
The stellar masses output by TNG are rescaled so as to match the known outcome of the TNG300 model at the superior resolution of the TNG100 simulation \citep[see][for more details]{2018MNRAS.475..648P}. Relative to TNG, the SC-SAM produces lower values of $m_*/M_{\rm halo}$ at halo masses above ($M_{\rm halo} \lesssim 10^{13} \ M_\odot/h$). In addition, the SC-SAM produces a sharper peak in the $m_*/M_{\rm halo}$ relation, and has considerably more scatter in this relation around its peak \citep[see also][]{Austen+2021}. As a result, the stellar mass (number density) selected SC SAM sample has a significantly larger number of galaxies occupying smaller haloes than the TNG sample.
%\rss{i did not understand the following comment} 
An additional effect we account for is the difference in the halo mass functions of the two finders \textsc{ROCKSTAR} (used in SC-SAM) and FoF (used in TNG), which affects the denominator of the $m_*/M_{\rm halo}$ relation. As seen in App.~\ref{app:halos}, \textsc{ROCKSTAR} finds 15\% more haloes at the mass scale of $M_{\rm halo} 10^{12} - 10^{13} \ M_\odot/h$ compared with FoF. To better compare the halo masses of the two catalogues, we abundance match the haloes in \textsc{ROCKSTAR} to those in FoF. We show the average stellar-to-halo mass ratio for the entire population of subhaloes and compare them against the result from the UniverseMachine (UM) Data Release 1 \citep{2019MNRAS.488.3143B}, which is a semi-empirical model that is designed to match the observed stellar mass function. We find that SC-SAM is in very good agreement with UM for large halo masses, whereas the rescaled TNG \citep[rTNG, see][for details]{2018MNRAS.475..648P} run provides a better match for intermediate halo masses around the peak in the stellar-to-halo mass relation. 
%rss stellar feedback not AGN FB. i prefer to discuss the reasons for the differences in a discussion section.
%The difference on small scales between SC-SAM and TNG is most likely due to a difference in the AGN feedback prescription. In particular, TNG ties the strength of this effect to the dispersion velocity, while in SC-SAM, it is tied to the maximum circular velocity of the halo at twice the NFW scale radius. Preliminary tests of the conjecture show that this change to the SC-SAM model indeed reduces the discrepancy between the two models; however, a more in-depth treatment is required before we can conclude whether this modification is sufficient to bring the two into agreement.

\iffalse
\begin{figure}
\centering  
\includegraphics[width=0.5\textwidth]{prof}
\caption{Satellite density profiles for a stellar-mass-selected sample with $\sim$3000 satellites at $z = 0$. In solid blue, we show the satellite profile of the SAM galaxy sample, while in solid red, we show the result for TNG. Stark cut}
\label{fig:prof}
\end{figure}https://www.overleaf.com/project/5f5d3d123f29aa0001f49bed
\fi

\subsection{Assembly bias parameters}
\label{sec:params}
In this section, we review the definitions of the various halo parameters employed in this study \citep[for more details, see][]{2020MNRAS.493.5506H}. App.~\ref{app:hab_corr} discusses the correlations between the various parameters in addition to the halo assembly bias measured for concentration and environment.

\subsubsection{Halo environment}
We adopt the following definition of halo environment to assess its effects on the large-scale galaxy distribution:
\begin{itemize}
\item[1.] Evaluate the dark matter overdensity field, $\delta (\mathbf{x})$, using cloud-in-cell (CIC) interpolation on a $256^3$ cubic lattice. Each cell has size of $205/256 \ {\rm Mpc}/h \approx 0.8 \ {\rm Mpc}/h$.
\item[2.] Smooth the overdensity field with a Gaussian kernel of smoothing scale $R_{\rm smooth} = 1.1 \ {\rm Mpc}/h$.
\item[3.] The halo local environment is thus determined by the value of the smoothed overdensity field in the cell where its centre-of-potential is located.
\end{itemize}
In previous works, we found that conditioning on this parameter in TNG leads to a substantial increase in the galaxy clustering on large scales \citep{2020MNRAS.493.5506H,2021MNRAS.501.1603H}. We conjectured that at a fixed mass, haloes residing in denser regions contain more galaxies on average than haloes in underdense regions due to experiencing more mergers and having a larger gas reservoir with which to form stars \citep{2007MNRAS.378..641A,2017A&A...598A.103P,2018MNRAS.476.5442P,2018MNRAS.473.2486S}. In App.~\ref{app:hab_corr}, we examine the correlation between environment and various other halo parameters. We find: on the high halo-mass end, the strongest correlation is between environment and the ratio $M_{\rm FoF}/M_{\rm 200}$; 
%\rss{what is M200 here? i don't think we have defined it}
on the low-mass end, it is between environment and velocity anisotropy, defined as $1-0.5\sigma_{\rm tan}/\sigma_{\rm rad}$, which were previously found to be good predictors of galaxy clustering. In Fig.~\ref{fig:hab}, we also present the halo clustering dependence on environment for both \textsc{ROCKSTAR} and FoF, confirming that haloes living in dense environments are substantially more clustered compared with haloes in underdense regions. This result is more noticeable for the \textsc{ROCKSTAR} haloes that tend to be better-deblended than the FoF ones, especially at these higher densities.

\subsubsection{Halo concentration}
The link between halo concentration and accretion history has been studied extensively in the literature \citep{1997ApJ...490..493N,Wechsler:2001cs,2014MNRAS.441..378L,2016MNRAS.460.1214L}. Very massive haloes with low concentration tend to have their subhaloes more spatially spread out as a result of having undergone a larger number of recent mergers \citep{sownak}. 
%\rss{i don't understand the following sentence. it seems like a more compact halo would lead to more clustered subhalos}
%\textbf{A richer subhalo structure might imply that these massive haloes of lower concentration have more highly clustered galaxies than more compact haloes at scales comparable to the halo virial radius. }
Therefore these halos may contribute to a boost in the number of objects, and hence clustering, at scales comparable to the virial radius of these halos. 
%rss the following is about galaxy formation and does not seem to fit here
%As we reach lower-mass haloes that only host 0 or 1 galaxies (i.e. only a central), we find more concentrated haloes are more likely to contain a central galaxy because their gravitational wells are deeper -- gas is more likely to collapse towards the centre and form stars. 
The effect of halo concentration on halo clustering and the relationship between halo concentration and halo mass have been thoroughly explored
\citep{2001MNRAS.321..559B,2014MNRAS.441..378L,2015ApJ...799..108D,2014MNRAS.441.3359D,2018MNRAS.474.5143M}. 

To obtain the halo concentration, we use the \textsc{ROCKSTAR} default definition, $c_{\rm NFW}$, which follows the Navarro-Frenk-White prescription \citep{1996ApJ...462..563N}, assuming that the density profiles of dark matter haloes are well-fitted by the formula
\begin{equation}    
    \rho_{\rm NFW}(r) = \frac{\rho_{\rm s}}{(r/r_{\rm s})\,(1+r/r_{\rm s})^{2}},
\end{equation}
where $\rho_{\rm s}$ is the characteristic density and $r_{\rm s}$ is the scale radius defined as
\begin{equation}     
    r_{\rm s} \equiv \frac{r_{\rm vir}}{c} \ ,
\end{equation}
with $r_{\rm vir}$ being the virial radius of the halo. We define this as the distance from the halo centre to the outer boundary within which the mean density is $\Delta_c$ times the critical density (see Section~\ref{sec:sam} for details on $\Delta_c$).

Fig.~\ref{fig:hab}, showing the halo assembly bias due to concentration, confirms the claim that less concentrated \textit{high-mass} haloes exhibit an excess in clustering, whereas less concentrated \textit{low-mass} haloes tend to be less clustered than their highly concentrated counterparts. Interestingly, we find very weak correlation between halo concentration and environment as defined in this study (see Fig.~\ref{fig:halo_corr}). We confirm the expected relationship between formation epoch and concentration.

\subsubsection{Halo spin}
Another halo parameter that we explore is halo spin, $\lambda$. This provides a measure of the angular momentum acquired by the halo. Its dependence on other halo properties, such as concentration and formation epoch, has been well studied \citep{2001ApJ...555..240B,2007MNRAS.376..215B,2016MNRAS.462..893R,2019MNRAS.486.1156J}. The measurement of this parameter is quite sensitive to the particle resolution \citep[the smaller the number of particles in a halo, the larger the error,][]{2014MNRAS.437.1894O,2017MNRAS.471.2871B}. In this work, we use the direct output from \textsc{ROCKSTAR}, which defines halo spin as \citet{1969ApJ...155..393P}:
\begin{equation}
    \lambda = \frac{J \sqrt{|E|}}{G M^{2.5}_{\rm vir}},
\end{equation}
where $J$ is the magnitude of the angular momentum of the halo and $E$ is the total energy of the halo. We show the correlation between spin and environment, and spin and concentration in Fig.~\ref{fig:halo_corr}.

\subsubsection{Properties computed from merger trees}
By computing select quantities directly from the \textsc{ROCKSTAR} merger trees and using the available bijective matches between \textsc{ROCKSTAR} and FoF, we can examine the effects in both. These quantities are listed as follows:

\begin{itemize}
    \item $z_{\rm form}$: the time at which the halo first acquires 50\% of its final mass. The characteristic formation epoch of a halo is a direct indicator of its past accretion history at fixed present-day mass. This quantity has a well-known connection to the halo concentration (which we recover in Fig.~\ref{fig:halo_corr}). Namely, the central regions of early-forming haloes collapse when the mean density of the Universe is higher and these halos are thus more concentrated.  
    \item $M_{\rm peak}$: the total mass of the particles within the virial radius of the halo (see Section~\ref{sec:sam} for definition of virial mass) at the time when it achieves its peak mass. A growing body of evidence seems to suggest that populating haloes based on their early-history properties (e.g. mass or velocity at time of infall or at their peak value) leads to a better agreement with observed galaxy clustering \citep[e.g.][]{2016MNRAS.460.3100C}. Haloes that have flown by a large object are well-known to be stripped of the outer layers of their dark matter component, retaining the baryonic core component. In such situations, the peak/infall mass acts as a better marker of their galaxy properties.
    \item $V_{\rm peak}$ the maximum value of $V_{\rm max}$ 
    %\rss{i found the z label to be confusing -- this is the maximum over radius}
    attained by the halo throughout its history, where $V_{\rm max}$ is the maximum value of the spherically averaged rotation curve for a given halo as a function of radius at some time epoch. Velocity-related parameters are thought to be more resilient to  well-known effects, such as tidal stripping, satellite disruption and gas expulsion (in the form of AGN feedback, supernova feedback, stellar winds, etc.), when compared with mass-related ones \citep[e.g.][]{2016MNRAS.459.3040G}. This suggests they are less prone to change during the galaxy formation process and therefore might offer a better proxy for galaxy occupation.
\end{itemize}

\section{Results}
\label{sec:res}
In this section, we present the main results of the study. In particular, we present comparisons between our mass-selected SC-SAM and TNG samples for various galaxy-halo statistical quantities, such as occupancy distributions, two-point clustering, the galaxy assembly bias signal, and additional probes describing the galaxy distribution. Furthermore, we explore the dependence of halo occupancy and gas content on the halo environment, studying the differences between the two models.

\subsection{Halo Occupation Distribution}
\label{sec:hod}
The occupancy of haloes at fixed mass, also known as HOD, is a standard statistical quantity used to examine the galaxy-halo connection. It provides information about the average number of galaxies as a function of halo mass and can be broken down into the respective contributions from central and satellite galaxies. Its shape, in the case of an LRG sample, has been studied extensively and is well approximated by the empirical formula given in \citet{Zheng:2004id}. One of the most common applications of HODs is in the construction of mock catalogues, i.e. galaxy catalogues meant to mimic the distribution of objects found in cosmological surveys. These mocks often assume a very simple relationship between haloes and the galaxies within them, motivated by the expected HOD of the population, in order to satisfy the hefty demands for volume and number of mock realizations \citep{2008RSPTA.366.4381B}.

\begin{figure*}
\centering  
\includegraphics[width=1\textwidth]{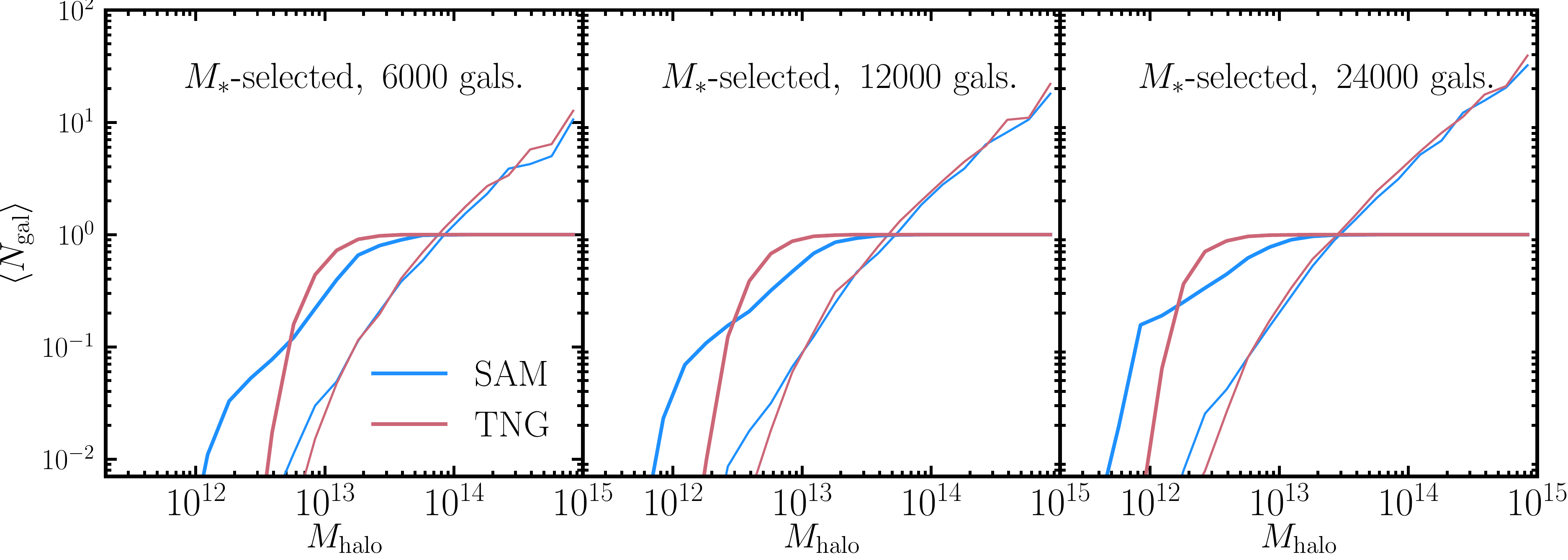}
\caption{Halo occupation distribution for a stellar-mass-selected sample with number densities of $n_{\rm gal} = 0.0005, \ 0.001, \ {\rm and} \ 0.002 \ [{\rm Mpc}/h]^{-3}$ (corresponding to 6000, 12000, and 24000 galaxies, respectively) at $z = 0$. In blue, we show the HOD of the SAM galaxy sample, while in red, we show the result for TNG. The galaxies are split into centrals (thick solid lines) and satellites (thin solid lines). The agreement between the TNG and SC-SAM satellite average occupation is very good for all panels. For all three number densities, the average HOD of the TNG centrals displays a sharper transition from 0 to 1 relative to the SC-SAM centrals. This is due to the tendency of SC-SAM to `paint' galaxies onto lower-mass haloes. The effect may be related to the particular implementation of AGN feedback or additional dependence of the halo gas content on parameters such as environment.}
\label{fig:hod_all}
\end{figure*}

\begin{figure*}
\centering  
\includegraphics[width=1\textwidth]{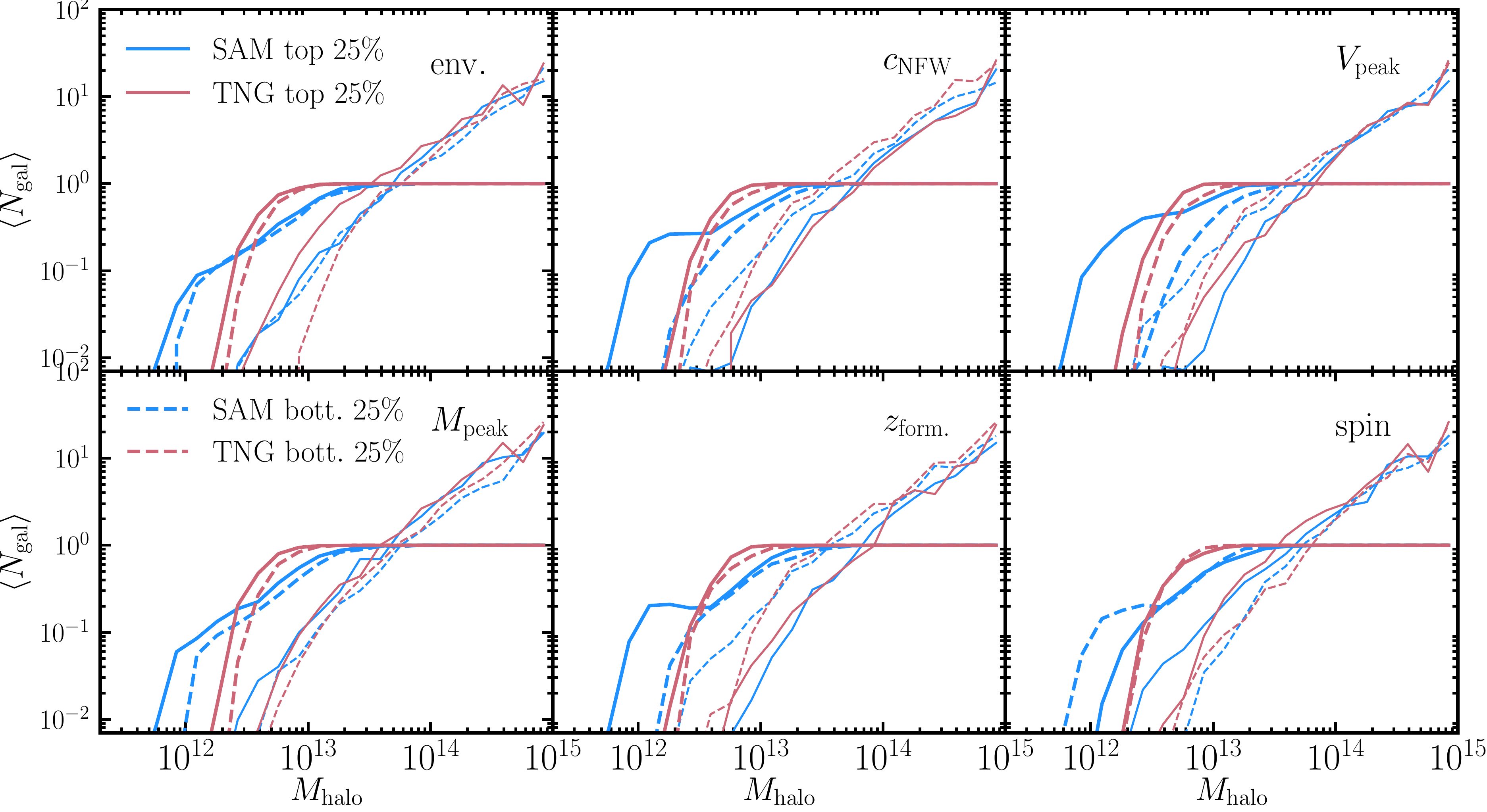}
\caption{
%\rss{i think the labels on the panels must be wrong: make sure they match the ones you have defined in the text}. 
Dependence on the halo parameters of the halo occupation distribution for a stellar-mass-selected sample with a galaxy number density of $n_{\rm gal} = 0.001 \ [{\rm Mpc}/h]^{-3}$ (corresponding to 12000 galaxies) at $z = 0$. In blue, we show the HOD of the SAM galaxy sample, while in red, we show the result for TNG. The galaxies are split into centrals (thick lines) and satellites (thin lines). In each mass bin, we select the top and bottom 25\% of haloes (solid and dashed lines, respectively) ordered by one of six properties: environment, concentration, formation epoch, spin, peak circular velocity, and peak mass (see Section~\ref{sec:params}). The qualitative response to the various parameters is similar in both models, with slight discrepancies. For example, the environmental effect is much stronger for TNG satellites and so is the response of the SC-SAM to
%\rss{picky thing: use consistent notation; should change to $V_{\rm peak}$ not $V_{\rm disk, peak}$ on plot label}
$V_{\rm peak}$, $z_{\rm form}$ and $c_{\rm NFW}$. The latter is likely due to the explicit appearance of these parameters in the model and/or the strong correlation between some of them (see Section~\ref{sec:sam} and App.~\ref{app:hab_corr}).}
\label{fig:hod_param_mstar}
\end{figure*}

In Fig.~\ref{fig:hod_all}, we show the halo occupation distribution as measured in the SAM and TNG, split into centrals and satellites for the three mass-selected samples considered, with number densities of $n_{\rm gal} = 0.0005, \ 0.001, \ {\rm and} \ 0.002 \ [{\rm Mpc}/h]^{-3}$. First focusing on the centrals, we see the SC-SAM sample has non-zero occupancy down to lower halo masses than TNG. Strikingly, the transition from 0 to 1 is markedly less steep in the case of the SC-SAM sample, where haloes are guaranteed a central (i.e. $\langle N_{\rm cen} \rangle \xrightarrow[]{}1$) only after they reach a minimum mass of $M_{\rm halo} \gtrsim 2 10^{13} \ M_\odot/h$. On the other hand, the average number of satellites per halo in the SAM displays very good agreement with TNG for all three samples. Although the HOD shape depends on the particular mass definitions adopted, changes to definitions only contribute a minor horizontal shift to the curves and do not alter the conclusions of our findings. Another factor is the different halo finding algorithms employed in both samples, but this effect is small and also does not affect our conclusions (see Fig.~\ref{fig:hod_rock})

Fig. \ref{fig:hod_param_mstar} demonstrates the dependence of the HOD on various halo parameters at $z = 0$ for a mass-selected sample with $n_{\rm gal} = 0.001 \ [{\rm Mpc}/h]^{-3}$, corresponding to a lower mass threshold of $M_{\rm gal} \approx 5 \times 10^{10} \ M_\odot/h$. 
%\rss{provide the corresponding mass limit here as well}
For each halo mass bin, we order the haloes based on a given property and then select the top and bottom 25\% of haloes based on this ranking. We compute the average occupation in each mass bin and show the results for the following 6 halo parameters: environment, concentration ($c_{\rm NFW}$), spin, peak circular velocity ($V_{\rm peak}$),
%rss changing to match definitions given above ($V_{\rm disk,peak}$), 
peak mass ($M_{\rm peak}$), and formation epoch ($z_{\rm form}$) (for more details, see Section~\ref{sec:params}). 
%\rss{use consistent terminology to the section where you defined these quantities}.
Qualitatively, the response for all of these parameters seems to be similar between the two models, confirming the findings of previous analyses, such as the typical inversion of the satellite and central occupation in the case of concentration and formation time. There are two noteworthy differences. The first is found in the case of the halo environment parameter, towards which the SC-SAM model appears to be less sensitive than TNG (especially for the satellites). The second is the stronger response to the following parameters in the case of SC-SAM: concentration, formation epoch, and to a lesser extent peak velocity. 
%rss again, i prefer to discuss this in a discussion section where one can explain it in more detail
%This difference is likely caused by parameters above appearing explicitly in the physics model of SC-SAM (see Section~\ref{sec:sam}), whereas the ab-initio TNG model relies on a smaller number of these ad-hoc relations. 
In the next section, we study how conditioning on each of these parameters manifests itself on the galaxy clustering.

\subsection{Galaxy clustering}
\label{sec:clust}
\begin{figure}
\centering  
\includegraphics[width=0.48\textwidth]{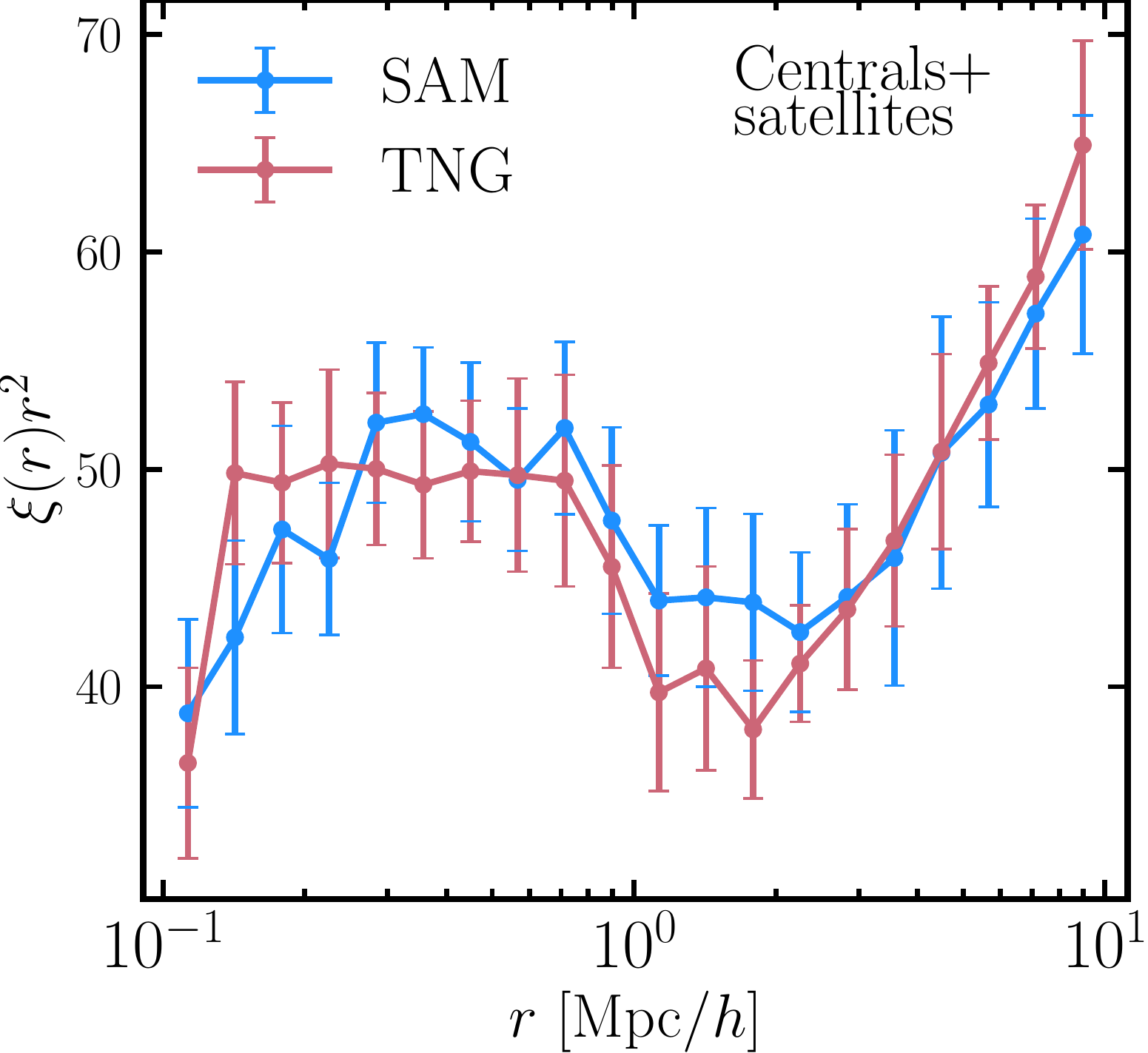} \\
\includegraphics[width=0.48\textwidth]{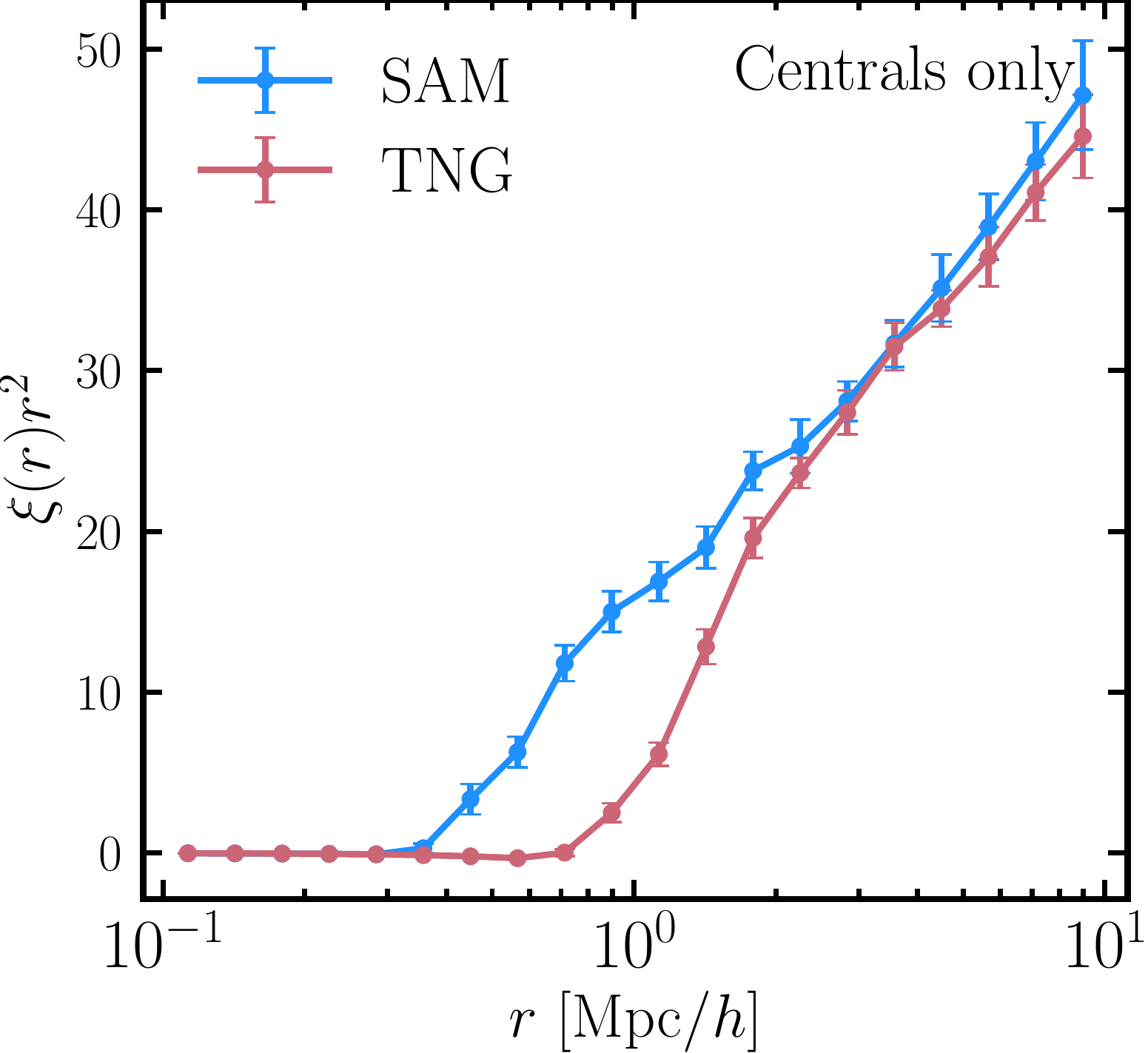} 
\caption{Correlation function for a stellar-mass-selected sample with a galaxy number density of $n_{\rm gal} = 0.001 \ [{\rm Mpc}/h]^{-3}$ (corresponding to 12000 galaxies) at $z = 0$. In blue, we show the two-point correlation function of the SAM galaxy sample, while in red, we show the result for TNG. The bottom panel shows the two-point clustering of only the centrals (roughly 9000 objects) for both samples. The error bars are computed using jackknifing. We see that on large scales, the agreement is excellent. On intermediate scales, centrals are slightly more clustered in SC-SAM, due to the tendency of SC-SAM to populate haloes of lower mass with smaller radii, thus reducing the range of the one-halo term. On small scales, SC-SAM exhibits higher clustering.}
\label{fig:corrfunc}
\end{figure}

The spatial two-point correlation function, $\xi(r)$, measures the excess probability of finding a pair of objects at a given separation, $r$, with respect to a random distribution. It is a vital tool in cosmology for exploring the three-dimensional distribution of any cosmological tracer and for constraining cosmological parameters. We compute the galaxy two-point correlation function using the natural estimator \citep{1980lssu.book.....P}:
\begin{equation}
    \hat \xi(r) = \frac{DD(r)}{RR(r)}-1 
\end{equation} 
via the package \textsc{Corrfunc} \citep{2020MNRAS.491.3022S}, assuming periodic boundary conditions in a box of size $L_{\rm box} = 205 \ {\rm Mpc}/h$. We estimate the uncertainties of the correlation function using jackknife resampling \citep{2009MNRAS.396...19N}. This is computed through dividing the simulation volume into 27 equally sized boxes before calculating the mean and jackknife errors on the correlation functions via the standard equations:
\begin{equation}
    {\bar \xi(r)}=\frac{1}{n}\sum_{i=1}^{n} {\xi}_i(r)
    \end{equation}
    \begin{equation}
    {\rm Var}[{\xi(r)}]=\frac{n-1}{n} \sum_{i=1}^{n} ({\xi_i(r)} - {\bar \xi(r)})^2 ,
\end{equation}
where $n = 27$ and ${\xi_i(r)}$ is the correlation function value at distance $r$ for subsample $i$ (i.e. excluding the galaxies residing within volume element $i$ in the correlation function computation). Additionally, we use jackknifing to estimate the error bars on ratios of correlation functions by first calculating the desired quantity in all $n_{\rm gal} = 0.001 \ [{\rm Mpc}/h]^{-3}$ subsamples, then computing the mean and standard deviation as shown above. In this way, we diminish some of the effects of sample variance and provide a way of quantifying the significance with which we detect deviations from unity.

In Fig.~\ref{fig:corrfunc}, we show the two-point correlation function, $\xi(r) \ r^2$, for the SAM and SC-SAM mass-selected samples at galaxy number density of $n_{\rm gal} = 0.001 \ [{\rm Mpc}/h]^{-3}$. In the upper panel, the two curves exhibit an excellent match on large scales (i.e. beyond the one-halo term, $\sim$2 Mpc/$h$); on intermediate scales ($\sim$1 Mpc/$h$) in the lower panel (and less so in the upper panel), the SC-SAM sample appears to be more strongly clustered. The behavior can be attributed to the following factors. First, SC-SAM tends to populate haloes of lower mass, which have smaller radii, so the exclusion radius, marking the transition between the one- and two-halo terms, manifests itself at smaller scales. The slightly stronger correlation signal at $r \sim$10 Mpc/$h$ of the SC-SAM centrals can be explained by the fact that SC-SAM preferentially populates highly concentrated haloes at fixed mass which at this mass-scale, exhibit strong halo assembly bias (see Fig.~\ref{fig:hab}). On small scales ($0.1 \ {\rm Mpc}/h \lesssim r \lesssim 1 \ {\rm Mpc}/h$), the satellite distribution of SC-SAM appears to be shifted to slightly larger scales. We conjecture that this is because the SAM satellites occupy slightly less concentrated haloes (see Fig.~\ref{fig:hod_all}) with a more spread-out satellite distribution, which shifts outward the one-halo clustering signal ($r \sim 1 \ {\rm Mpc}/h$). Another noticeable feature is the larger size of the error bars on the upper panel as a result of the larger shot noise of the satellites compared with the centrals. The error bars indicate the amount of variance resulting from the limited simulation volume, but the relative agreement or disagreement of the two models cannot be interpreted based on those, as the SAM and TNG results are calculated based on the same underlying halo distribution.

\subsection{Galaxy assembly bias signature}
\label{sec:gab}
Galaxy assembly bias can be thought of as a ``convolution'' between two separate effects. It is a result of halo assembly bias and occupancy variation, where the former is the dependence of the clustering of DM haloes on parameters other than mass,  %\citep{2007MNRAS.374.1303C}, 
while the latter refers to the dependency of halo occupancy on halo parameters other than mass. A standard way to measure galaxy assembly bias is by comparing the two point-correlation function from a given sample with a sample in which the galaxies populating haloes are randomly reassigned to haloes within the same mass bin \citep{2007MNRAS.374.1303C}. This ``shuffling'' eliminates the dependence of halo occupancy on any secondary properties other than halo mass. If the ratio of the shuffled to unshuffled sample differs from zero, then the galaxy assembly bias signal is non-zero. This method of defining galaxy assembly bias is only meaningfully measurable on large scales. Therefore, we opt to preserve the relative positions of galaxies with respect to their host halo centre when transferring them to their newly assigned host. In this way, the shuffled sample maintains the same one-halo term as the original sample, and the comparison is purely between the two-halo term contribution to both.

\begin{figure*}
\centering  
\includegraphics[width=1\textwidth]{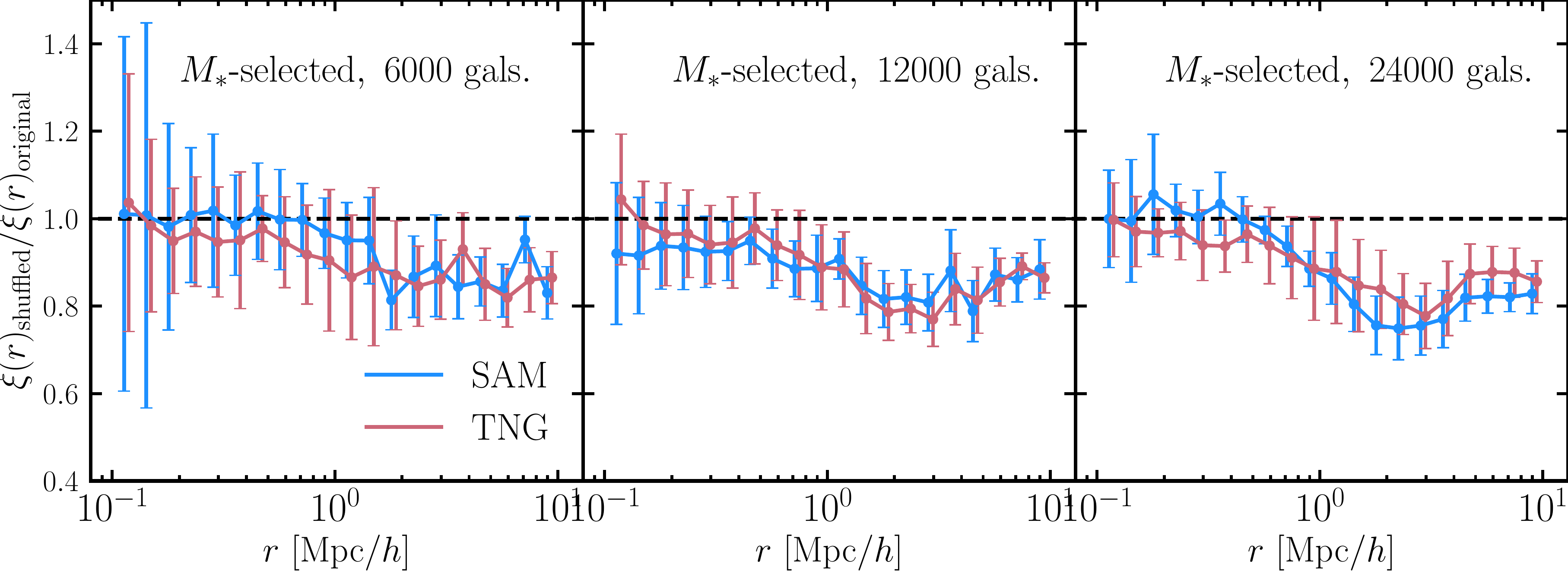}
\caption{Galaxy assembly bias of a stellar-mass-selected sample with number densities of $n_{\rm gal} = 0.0005, \ 0.001, \ {\rm and} \ 0.002 \ [{\rm Mpc}/h]^{-3}$ (corresponding to 6000, 12000, and 24000 galaxies, respectively) at $z = 0$. The signal for each model is computed by randomly shuffling the halo occupations at fixed halo mass, while preserving the one-halo term. In blue, we show the galaxy assembly bias signal of the SAM galaxy sample, while in red, we show the result for TNG. The error bars are computed using jackknifing of the ratios. A deviation from unity suggests that a mass-only HOD is incapable of matching the clustering of LRG-like galaxies. The signal is very similar for both models in the first two panels, but shows a slightly larger discrepancy in the third, where the SC-SAM galaxies display even stronger galaxy assembly bias.}
\label{fig:shuffle_all}
\end{figure*}

\begin{figure*}
\centering  
\includegraphics[width=1\textwidth]{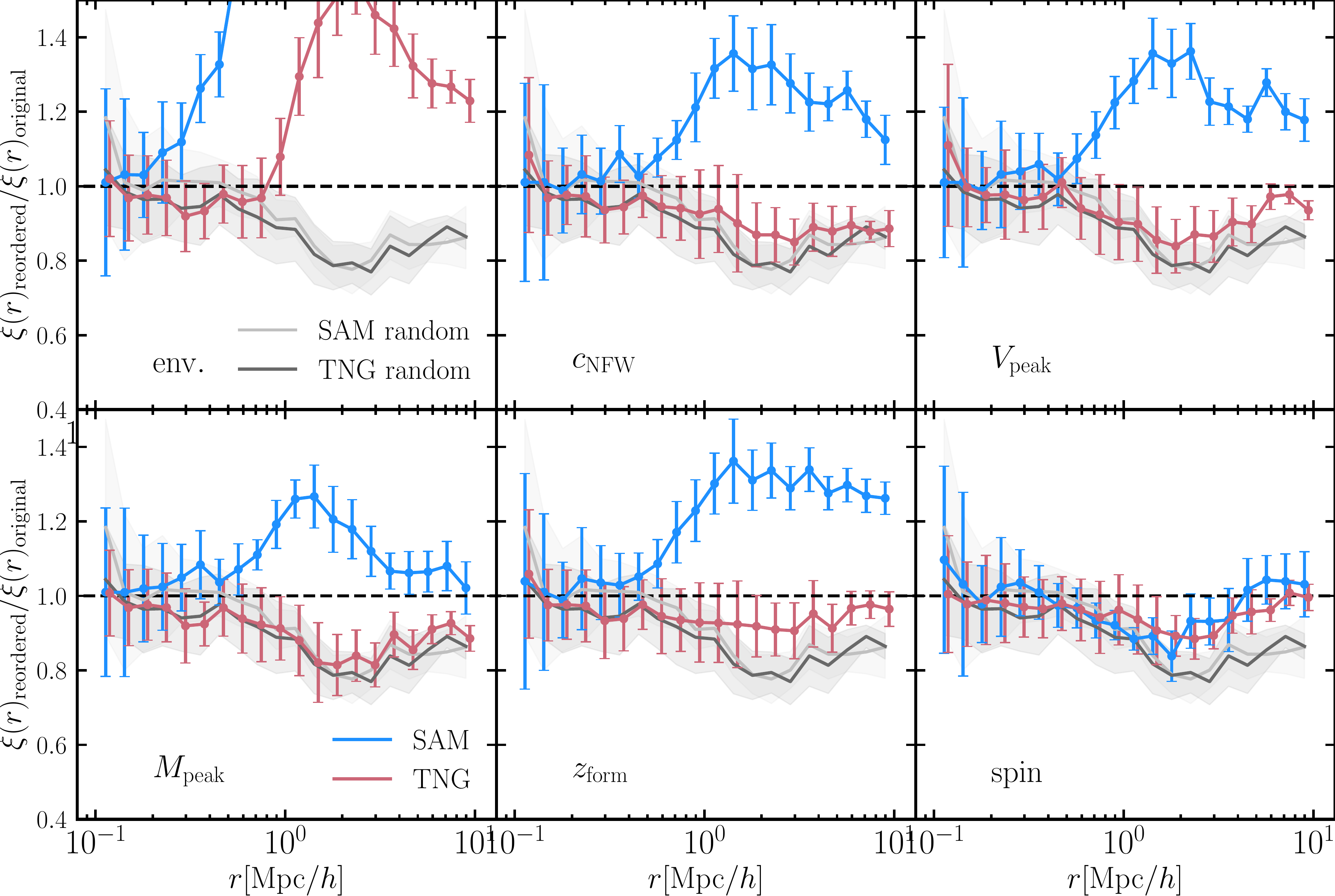}
\caption{Dependence on different halo parameters of the galaxy assembly bias for a stellar-mass-selected sample with a galaxy number density of $n_{\rm gal} = 0.001 \ [{\rm Mpc}/h]^{-3}$ (corresponding to 12000 galaxies) at $z = 0$. In blue, we show the galaxy assembly bias signal of the SAM galaxy sample, while in red, we show the result for TNG. In gray and black, we show the SC-SAM and TNG curves from the middle panel of Fig.~\ref{fig:shuffle_all}. Error bars are computed using jackknifing. The environmental dependence, which is significant in both models, is stronger in the SAM. This is likely because the halo assembly bias with respect to environment is stronger for the \textsc{ROCKSTAR} catalog compared with the FoF catalogue. As seen in Fig.~\ref{fig:hod_param_mstar}, we see that the SC-SAM sample is very sensitive to the parameters: concentration, formation epoch, and peak circular velocity, all of which appear explicitly in the SAM model (see Section~\ref{sec:sam}).}
\label{fig:shuffle_param_mstar}
\end{figure*}

In Fig.~\ref{fig:shuffle_all}, we illustrate the galaxy assembly bias signal for the three mass-selected samples considered in this study, with number densities of $n_{\rm gal} = 0.0005, \ 0.001, \ {\rm and} \ 0.002 \ [{\rm Mpc}/h]^{-3}$. We shuffle the halo occupation number in mass bins of width $\Delta \log(M_{\rm halo}) = 0.2 \ {\rm dex}$. For $r \gtrsim 1 \ {\rm Mpc}/h$, we find that the ratio deviates on average by about 15\%, 18\%, and 20\% for the three samples, respectively. Moreover, it is noteworthy that the SC-SAM and TNG samples appear in very good agreement with each other, with the largest difference between the two being found in the sample with highest number density. On small scales ($r \lesssim 1 \ {\rm Mpc}/h$), the ratio is consistent with one, as we preserve the one-halo term when performing the shuffling. This finding suggests the SC-SAM model naturally induces a non-negligible galaxy assembly bias signal, which is similar to that seen in a hydrodynamical simulation. 

Studying the response of both models to various halo properties is important for understanding the source of the assembly bias signal. Instead of randomly shuffling the halo occupancies in bins of halo mass, we now sort the haloes by a selected property: environment, concentration, peak velocity, peak mass, formation epoch and spin, in either descending or ascending order, and then distribute the galaxy occupations to the sorted haloes in descending order. The result of this exercise is shown in Fig.~\ref{fig:shuffle_param_mstar}. The strongest response of the clustering is to halo environment. We conjecture that it is predominantly the result of halo assembly bias, which explains the stronger signal for the SC-SAM sample. On one hand, as a result of the flatter HOD shape of the SC-SAM centrals (see Fig.~\ref{fig:hod_all}), the reordering procedure employed when measuring galaxy assembly bias preferably picks haloes in dense environments across a wider mass scale compared with TNG. This inevitably boosts the clustering signal as these haloes are significantly more biased. To test this hypothesis, we create an ad-hoc SAM galaxy sample assuming the HOD shape of TNG and find that the assembly bias signature becomes nearly identical to that of TNG with differences comparable to the halo assembly bias signal seen in Fig.~\ref{fig:hab}. On the other hand, a secondary effect comes from the fact that SC-SAM uses the \textsc{ROCKSTAR} halo catalogue which exhibits a slightly stronger dependence on environment than FoF (see Fig.~\ref{fig:hab}). 
%\rss{hmm. i'm not sure the differences seen in fig. B1 are large enough to explain this very large difference. it would be good to check what you get if you use the rockstar halos for the TNG analysis}
Occupancy variation also plays a role in determining galaxy assembly bias: the response to environment is similar in both, with TNG exhibiting a slightly stronger dependence (see Fig.~\ref{fig:hod_param_mstar}). 
%\rss{here, the SAM and TNG show a similar level of dependence on environment, or TNG is perhaps even a bit stronger}
Another interesting observation, which echoes our finding from Fig.~\ref{fig:hod_param_mstar}, is that reordering the haloes in terms of their concentration, circular velocity, formation epoch, and peak mass results in a significant increase in the SC-SAM clustering, whereas TNG is less sensitive to these parameters (a similar conclusion was reached in \citet{2020MNRAS.493.5506H}). This makes sense in light of the similarly stronger dependence of the halo occupancy on these parameters in the SAM.
%An explanation for this behavior in the SAM would be the explicit appearance of these parameters in the relations that determine the amount of injected gas, cooling and star formation, and the strength of the feedback processes (see Section \ref{sec:sam}). An exception is peak mass, as it does not appear explicitly; however, it is strongly correlated with the circular velocity (see Fig.~\ref{fig:halo_corr}).

\subsection{Galaxies and their gaseous environment}
\label{sec:env}
One of the major differences between SAMs and numerical hydrodynamic simulations is that the numerical simulations explicitly simulate the distribution of gas on both large and small scales, and can therefore make detailed predictions about the temperature and density of this gas, as well as how these properties are affected by the baryonic feedback processes such as stellar and AGN feedback. SAMs make simplifying assumptions, such as that halo gas is in an isothermal sphere with a fixed density profile, and cannot make any predictions for how gas is distributed outside of halos. The availability of gas is expected to have an important affect on observable galaxy properties.  In order to interpret our results, it is therefore interesting to explore whether the way in which gas traces the underlying dark matter shows any dependence on environment, and whether this is the same in the two models. 

%Semi-analytic models and hydrodynamical simulations both implement ``sub-grid'' recipes to emulate processes involved in galaxy formation that cannot be resolved and simulated directly. These processes include star formation, stellar feedback, and black hole formation, growth, and feedback. An advantage of simulating hydrodynamics over SAMs is that one traces the gas particles directly before triggering the stochastic processes (mentioned above) rather than injecting them based on a handful of halo properties (see Section~\ref{sec:sam}). This is an important distinction: the distribution of the condensing gas determines the location of galaxies, which is our most direct observable in galaxy surveys. If the secondary parameters involved in controlling the gas content in SAMs do not reflect those in hydrodynamical simulations, this may result in a vastly different galaxy distribution.

\begin{figure}
\centering  
\includegraphics[width=0.48\textwidth]{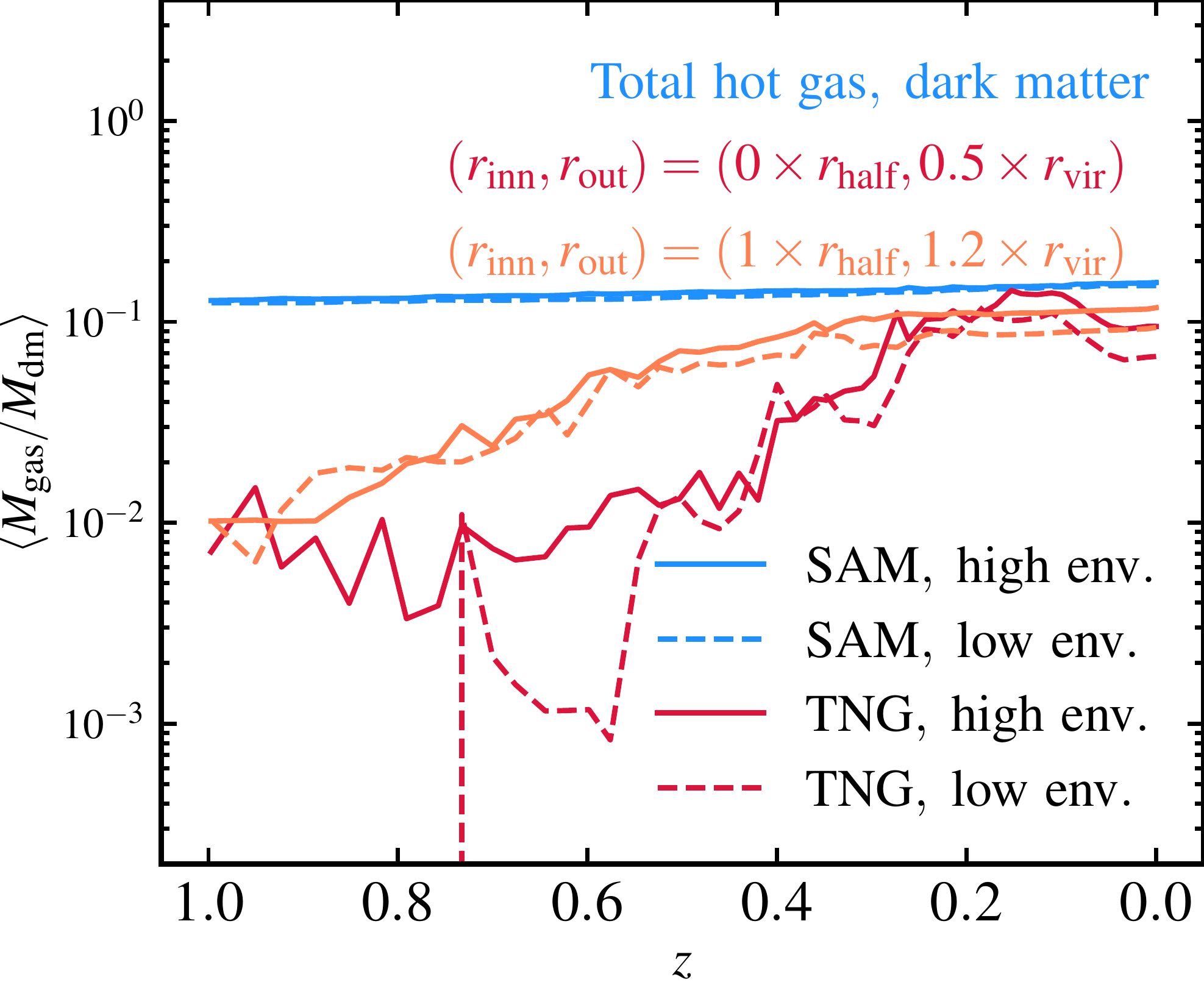}
\caption{Ratio of the gas to dark matter mass for SC-SAM (blue) and TNG (red). In the case of TNG, these numbers come from counting the gas particles contained within an annulus with inner and outer radius pairs -  $(0, 1/2 \times r_{\rm vir})$ and $(r_{\rm half}, 1.2 \times r_{\rm vir})$ - as a function of redshift for $\log M = 13$ haloes belonging to high- (marked with solid lines) and low- (marked with dashed lines) density environments. In the case of SC-SAM, we take the ratio between the total amount of hot gas and the total dark matter mass, as the model does not output the individual gas profiles of haloes. We see that the gas content in TNG haloes exhibits a much stronger connection with environment; the effect is minimal in SC-SAM. In TNG, gas is supplied to the halo through filaments and sheets and is thus inherently environment-dependent; in SC-SAM, the amount of gas injected into and ejected from the halo is determined by other halo properties (halo mass, concentration, maximum circular velocity).}
\label{fig:gas_content_env}
\end{figure}

One might intuitively expect to find that objects lying in high-density regions have a higher gas content and vice versa for haloes in low-density environments. To test this conjecture, we show the ratio between gas and dark-matter content in TNG and SAM haloes as a function of redshift in Fig.~\ref{fig:gas_content_env}. We select abundance-matched haloes with $\log M_{\rm halo} = 13.0 - 13.1$ from the two catalogues. We obtain the abundance matched samples by rank-ordering the haloes by halo mass in both catalogues. Then we find the halo indices corresponding to the mass range of interest in the FoF catalogue and select the haloes from the ROCKSTAR catalogue based on those.

%\rss{i'm not sure what "abundance matched" halos means if you are just selecting halos in a given mass range.}
This scale roughly corresponds to the average host halo mass for the stellar-mass-selected galaxy sample presented in this study. We then chose the top and bottom 10\% of haloes based on the value of their environment parameter, resulting in four sets of 50 haloes each. In the case of SC-SAM, we do not resolve the individual gas particles, so the ratio between the total amount of gas in the halo and its virial dark-matter mass is shown instead. In the case of TNG, we further show the differential gas mass within shells surrounding the central galaxy. We adopt the following inner and outer radii pairs: $(r_{\rm inner}, r_{\rm outer}) = \{(0, 1/2 \times r_{\rm vir}), (r_{\rm half}, 1.2 \times r_{\rm vir})\}$, where $r_{\rm half}$ is the stellar half-mass radius for the central subhalo reported in TNG and $r_{\rm vir}$ is the virial radius (`Group\_R\_TopHat200'). 
%\rss{how do these compare to the halo virial radius? my rough rule of thumb is that rgal is roughly rvir/50. this would seem to imply that you are going way beyond the virial radius, but am i missing something? to compare with the SAM, it would be good to have a measurement within the rockstar halo rvir}.
These are chosen to be at the scale of the typical galaxy formation processes, to get a sense of the gas reservoir available locally for forming stars and accreting onto the central galaxy. We see that the SAM result appears to be almost completely insensitive to environmental variations.
%, with the strongest effects seen at $z \approx 0.5$. 
On the other hand, the amount of gas surrounding the TNG centrals exhibits a much more noticeable dependence on environment. We emphasize that our halo selection and environment are defined at $z = 0$. We can connect the findings of this section to some of the takeaways from Fig.~\ref{fig:hod_all}: namely, we noted that the dependence of the average number of satellites per halo mass on environment was much weaker in SC-SAM compared with TNG. This is in agreement with the conjecture of Fig.~\ref{fig:gas_content_env} that haloes in higher-density regions host a larger number of satellites due to their ex-situ introduction into the main halo through filaments (along with gas).
%and in a small fraction of the cases, in-situ star-burst formation as a result of halo mergers. 
There are additional environmental effects that go in the opposite direction of destroying satellites in dense environments such as ram pressure stripping and tidal interactions with the central, but these effects appear to be of secondary importance, as they do not have a visibly strong impact on the satellite occupancy (see Fig.~\ref{fig:hod_all}) and quantifying their size is not the subject of the current study.
%\rss{not sure about this interpretation. first, are the SAM results shown in Fig. 3 for the "raw" SAM or are they affected by the selection that forces the same satellite fraction in the SAM and TNG? second, if the TNG measurements are really for the gas out to way beyond the virial radius, i'm not sure this would be expected to have a direct impact on the satellites. i would expect a stronger affect on the central, but we see almost no difference in the HOD with environment for TNG for central. in fact, more gas might lead to more ram pressure stripping of satellites and so lower stellar masses in them! so, i'm puzzled about how to tie this all together. }
We make available time-lapse visualisations of the selected TNG haloes (the entire FoF group) here: \href{https://drive.google.com/drive/folders/1AZX8eecrKwHGXtqGSf5XioXhI5kFkswI?usp=sharing}{here}.

\begin{figure}
\centering  
\includegraphics[width=0.48\textwidth]{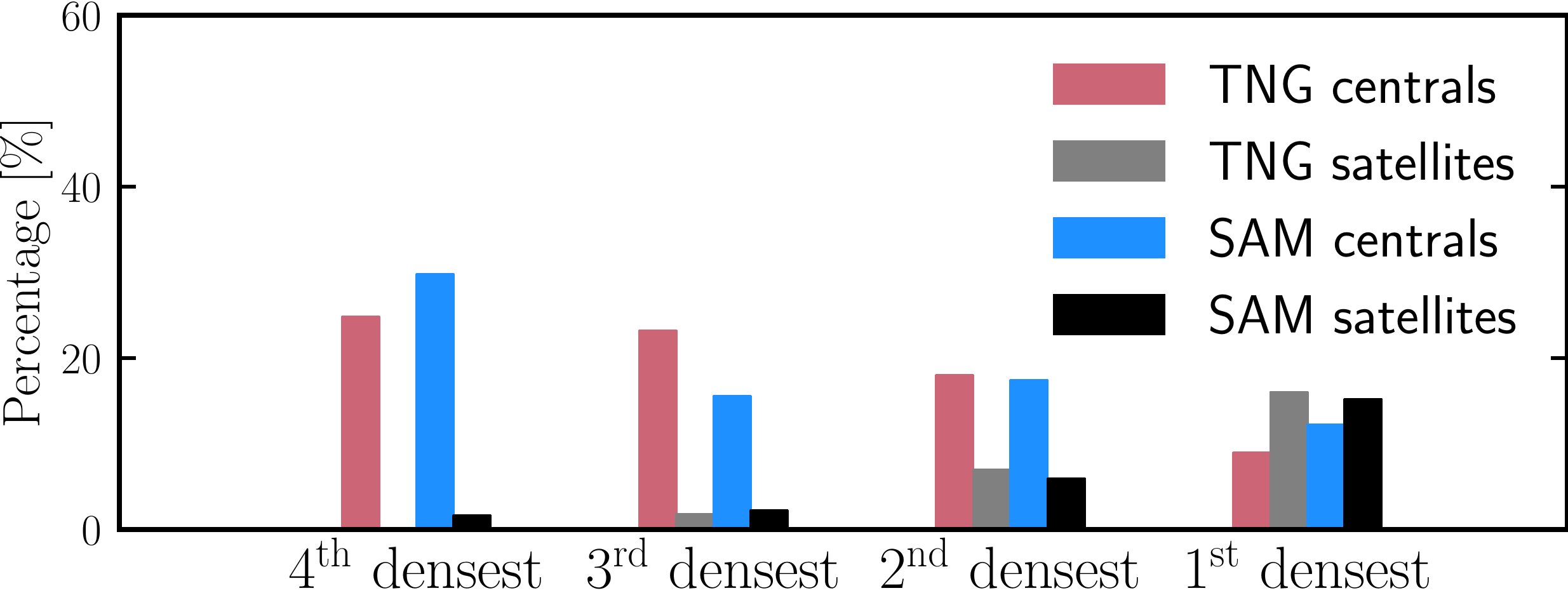}
\caption{Percentage of galaxies found in the different density regions for SAM and TNG, split into centrals and satellites. The sum of the central and satellite fractions for each model (TNG and SC-SAM) in each of the four density regions contributes 25\%.}
\label{fig:step}
\end{figure}

Another feature of Fig.~\ref{fig:hod_all} that might be affected by the absence of environmental dependence in the SAM is the central occupation. To test this, we examine the spatial distribution of the galaxies as a function of local environment. We order the galaxies in the TNG stellar-mass-selected sample by their local environment and split the sample into 4 subsamples of equal size (i.e. 3000 galaxies in each). Using the same threshold values of the local environment, we obtain 4 subsamples for SC-SAM and display the results in Fig. \ref{fig:step}, separated into centrals and satellites.
%\rss{it is almost always easier to digest information that is in a figure rather than a table -- could you visualize this table by just plotting the fractions arrayed from left to right from least to most dense? or perhaps even plot the average density in each of the four quadrants on the x-axis?}
The most conspicuous discrepancy between the two models is in the densest and least dense regions (which have the most pronounced effect on the clustering because their galaxy bias displays the largest fluctuation). There is a substantially larger number of SC-SAM centrals (29.8\%) and satellites (1.6\%) inhabiting the least dense region in contrast to TNG (24.8\% and 0.2\%) and a somewhat larger number of SC-SAM centrals (12.2\%) residing in the densest regions compared with TNG (9.0\%). The third densest region compensates for the difference in central fractions with TNG containing 23.2\% of them while SC-SAM only 15.6\%.
%\rss{this does not match the numbers in the table, unless i'm misreading something}.
The larger number of SAM galaxies in the least dense region seems consistent with the tendency seen in Fig.~\ref{fig:hod_all} of the SAM to populate haloes of lower masses with galaxies. This effect, in conjunction with the larger fraction of SAM centrals in the densest region (with the highest bias), yields a consistent correlation function between the two samples (see Fig.~\ref{fig:corrfunc}). 
%The findings of this section suggest that galaxy models can be differentiated and better calibrated by using the two-point clustering of various tracers. 
%rss move to discussion/interpretation
%Incorporating an explicit dependence on environment in SC-SAM would likely increase the similarity between TNG and SC-SAM. A more sophisticated treatment of the satellite galaxies in SC-SAM could also take into account the gradual removal of gas due to ram pressure stripping in high-density environments. 

\iffalse
\begin{table}
        \begin{center}
          \begin{tabular}{| c | c | c | c | c |}
          \hline\hline
          Galaxy type & 4th densest & 3rd densest & 2nd densest & 1st densest \\ [0.5ex]
          \hline
          SAM centrals & 29.783 & 15.583 & 17.400 & 12.233 \\ [1ex]
          SAM satellites & 1.633 & 2.217 & 5.967 & 15.183 \\ [1ex]
          \hline
          TNG centrals & 24.800 & 23.192 & 18.025 & 8.983 \\ [1ex]
          TNG satellites & 0.200 & 1.808 & 6.975 & 16.017 \\ [1ex]
          \hline
          \hline
          \end{tabular}
        \end{center}
        \caption{Percentage of galaxies found in the different density regions for SAM and TNG.}
        \label{tab:sims}
\end{table}
\fi

\subsection{Bias and correlation coefficient}
\label{sec:bias}
%A large fraction of the information on the matter distribution is encoded in the two-point clustering or in the power spectrum of the matter density fluctuations as a function of scale and redshift. However, since we cannot observe all matter in the Universe directly, we need to rely on tracers of the large-scale structure. In cosmological surveys, we use galaxies to trace the underlying mass distribution; therefore, it is crucial to understand the relationship between the large-scale distribution of matter and galaxies. 
In traditional cosmological studies of the matter distribution and its relationship to cosmological parameters, it is common to describe the relationship between observable tracers (galaxies) and the underlying mass distribution in terms of a ``bias" parameter. 
The galaxy auto-correlation function, $\xi_{gg}(r)$ is related to the matter correlation function, $\xi_{mm}(r)$, through the real-space galaxy bias, $\tilde b (r)$, in the following way:
\begin{equation}
    \xi_{gg}(r) = \tilde b^2(r) \xi_{mm}(r).
\end{equation}
Furthermore, one can study the galaxy-matter connection through their cross-correlation function, $\xi_{gm}(r)$, which can be related to $\xi_{mm}$ through $\tilde b$
and the cross-correlation coefficient, 
$\tilde r$ \citep{2008MNRAS.388....2H,2018PhR...733....1D}:
\begin{equation}
    \xi_{gm}(r) = \tilde b(r) \tilde r(r) \xi_{mm}(r),
\end{equation}
where the cross-correlation coefficient is given by:
\begin{equation}
    \tilde r(r) = \frac{\xi_{gm}(r)}{[\xi_{gg}(r) \ \xi_{mm}(r)]^{1/2}}.
\end{equation} 
The equations above are general and may be taken as definitions of the scale-dependent galaxy bias $\tilde b(r)$ and cross-correlation coefficient $\tilde r(r)$, although alternative definitions exist and are not equivalent. We note that the quantity $\tilde r(r)$ in real space is not constrained to be less than or equal to one, as is the case in Fourier space. However, one expects $\tilde r(r)$ to approach unity on large scales,  where the observed correlation is sourced from the gravitational field of the total matter. 

\begin{figure}
\centering  
\includegraphics[width=0.48\textwidth]{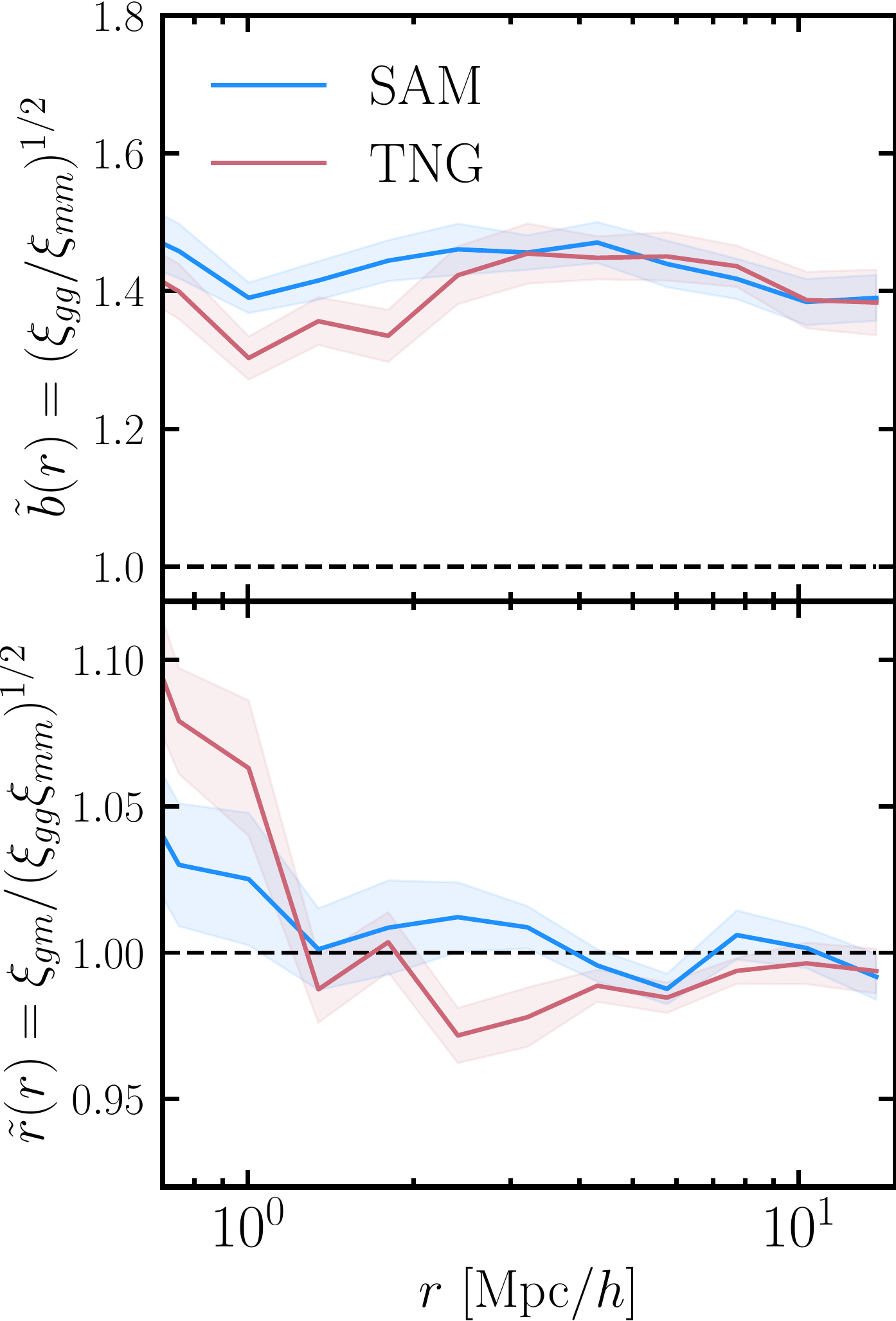}
\caption{Bias and cross-correlation coefficient for a stellar-mass-selected sample with a galaxy number density of $n_{\rm gal} = 0.001 \ [{\rm Mpc}/h]^{-3}$ (corresponding to 12000 galaxies) at $z = 0$. In blue, we show the bias and correlation coefficient of the SAM galaxy sample, while in red, we show the result for TNG. Error bars are derived using jackknifing. The bias in both models show good agreement and approaches $\tilde b(r) \approx 1.4$ on large scales. The cross-correlation coefficient for TNG is lower on intermediate scales, suggesting the linear bias assumption is imperfect. In the case of SC-SAM, it is consistent with one.}
\label{fig:bias_corr_coeff}
\end{figure}

Fig.~\ref{fig:bias_corr_coeff} shows the galaxy bias and the cross-correlation coefficient for a stellar-mass-selected sample with number density of $n_{\rm gal} = 0.001 \ [{\rm Mpc}/h]^{-3}$ for SC-SAM and TNG. The galaxy bias is consistent between the two models and levels out to about $\tilde b(r) \approx 1.4$ around $\sim$10 Mpc/$h$ at the onset of the linear bias regime. This result is expected, as the galaxy clustering matches well (Fig.~\ref{fig:corrfunc}) and the matter-matter correlation function is the same for both, being provided by dark matter particles in the simulation. The cross-correlation coefficient approaches one for $r \sim 1 \ {\rm Mpc}/h$, and on smaller scales, the two models show a moderate difference. In particular, for TNG, we find that $\tilde r(r) \approx 0.97$ for $r \approx 3 \ {\rm Mpc}/h$, suggesting that the linear bias approximation breaks down on these scales. On the other hand, the cross-correlation coefficient of the SC-SAM sample is consistent with one.

\subsection{Additional probes of the galaxy distribution}
\label{sec:stats}
In this section, we explore alternative statistical probes of the large-scale galaxy distribution, such as weak lensing, galaxy-void cross-correlation, and cumulants of the smoothed galaxy density field. We pursue the goal of further describing the galaxy distribution of the TNG and SC-SAM samples in addition to documenting the differences between the two models.

\subsubsection{Density field cumulants}

The cumulants of the density field are measured from the moments of the smoothed galaxy overdensity field. They can be understood as degenerate $N$-point correlation functions or as integrated monopole moments of the bispectrum, which are closely related to neighbour counts both in their physical interpretation as well as in their algorithmic implementation. Of particular importance is the scale of galaxy clusters ($\sim$3 Mpc/$h$ to 10 Mpc/$h$), where we expect that galaxy population methods could exhibit substantial differences \citep{1994A&A...291..697B,1994MNRAS.268..913G}. The procedure we follow (similar to \citet{2021MNRAS.501.1603H}) can be outlined as follows:
\begin{itemize}
\item Divide the TNG box into $512^3$ cubes  of side $\sim 0.4$ Mpc$/h$ and compute the counts-in-cell density field in each as $\delta_i = N_i/{\bar N}-1$.
\item Convolve it with a Gaussian filter of smoothing scale $R = \{3.75, 5, 6.25, 7.5, 8.75, 10\}$ Mpc/$h$ to get the smoothed density field $\delta_R$.
\item Compute the third moment of the density contrast by averaging the cubed overdensity over all cells, $\langle \delta^3_R \rangle$.
\item Study the change of the third moment values as a function of smoothing scale for different galaxy distributions.
\end{itemize}

\begin{figure}
\centering  
\includegraphics[width=0.48\textwidth]{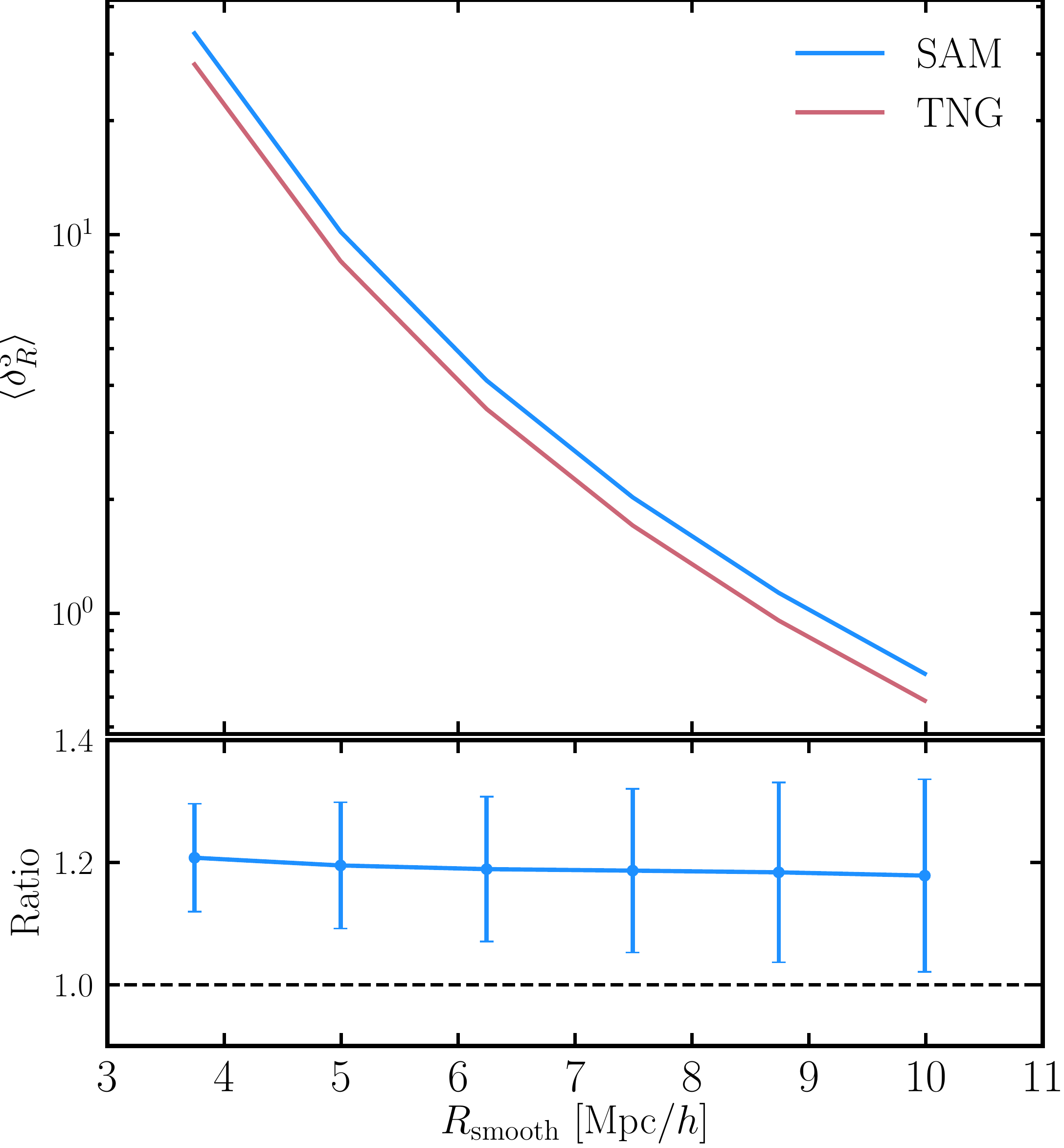}
\caption{Cumulant of the smoothed galaxy density, $\langle \delta^3_R \rangle$, as a function of smoothing scale for a stellar-mass-selected sample with a galaxy number density of $n_{\rm gal} = 0.001 \ [{\rm Mpc}/h]^{-3}$ (corresponding to 12000 galaxies) at $z = 0$. In blue, we show the cumulative density of the SAM galaxy sample, while in red, we show the result for TNG. $\langle \delta^3_R \rangle$ decreases as we increase the smoothing. The lower panel, showing the ratio between the two curves, presents a $\sim$20\% offset between the two models, despite the well-matched two-point clustering of the samples. This suggests that including higher-point statistics in the analysis and calibration of galaxy formation models can help us differentiate between them and allows us to obtain more accurate depictions of the large-scale galaxy distribution.
%\rss{compress the y-axis of the bottom plot to better illustrate the data.} 
}
\label{fig:third}
\end{figure}

In Fig.~\ref{fig:corrfunc}, we show that the two-point correlation functions of the SC-SAM and TNG galaxy populations exhibit excellent agreement with each other. However, this does not guarantee the consistency of higher-order statistics. The averaged third moment of the galaxy density field, $\langle \delta^3_R \rangle$, displayed in Fig.~\ref{fig:third}, offers a compressed view of the three-point correlation function. The upper panel shows  $\langle \delta^3_R \rangle$ as a function of smoothing scale. Increasing the smoothing scale $R$ smooths out more and more fluctuations of the galaxy density field; thus, the value of $\langle \delta^3_R \rangle$ decreases as expected. The lower panel shows the ratio of the SC-SAM curve to TNG. Surprisingly, we find that  $\langle \delta^3_R \rangle$ differs by $\sim$20\% despite the near perfect match of the two-point correlation functions. This finding argues strongly in favor of including alternative statistical probes when we compare and calibrate galaxy population models. Recovering the two-point correlation function may not be a sufficient condition for generating realistic galaxy catalogues.

\subsubsection{Galaxy-galaxy lensing}

Galaxy-galaxy lensing, also known as stacked lensing, offers a unique window for mapping the underlying matter distribution by measuring the cross-correlation of large-scale structure tracers with the shapes of background galaxies. Stacked lensing measurements are expected to be one of the most powerful probes of modern cosmology, allowing cosmologists to address fundamental physics questions such as the nature of dark energy and dark matter and the mass hierarchy of neutrinos \citep[e.g.][]{2011PhRvD..83b3008O}. Furthermore, by combining stacked lensing and auto-correlation measures of the same foreground galaxies, one can constrain cosmology by breaking degeneracies between galaxy bias and cosmological parameters \citep[e.g.][]{2005PhRvD..71d3511S,2020arXiv200203867S}. Hydrodynamical simulations are invaluable in this manner: they allow us to compare the weak lensing signal, as predicted by various galaxy population prescriptions, and evaluate their ability to match the ``true'' galaxy distribution, where ``true'' is defined by the hydro simulation. Although the hydro model may not be fully consistent with observations, it is still a plausible model for how matter, baryonic and dark, is distributed in the Universe. It is, therefore, crucial to demonstrate that the galaxy population techniques we adopt when creating high-fidelity mock catalogues are flexible enough to recover various statistical properties of the galaxy sample in a hydro simulation.

\begin{figure}
\centering  
\includegraphics[width=0.48\textwidth]{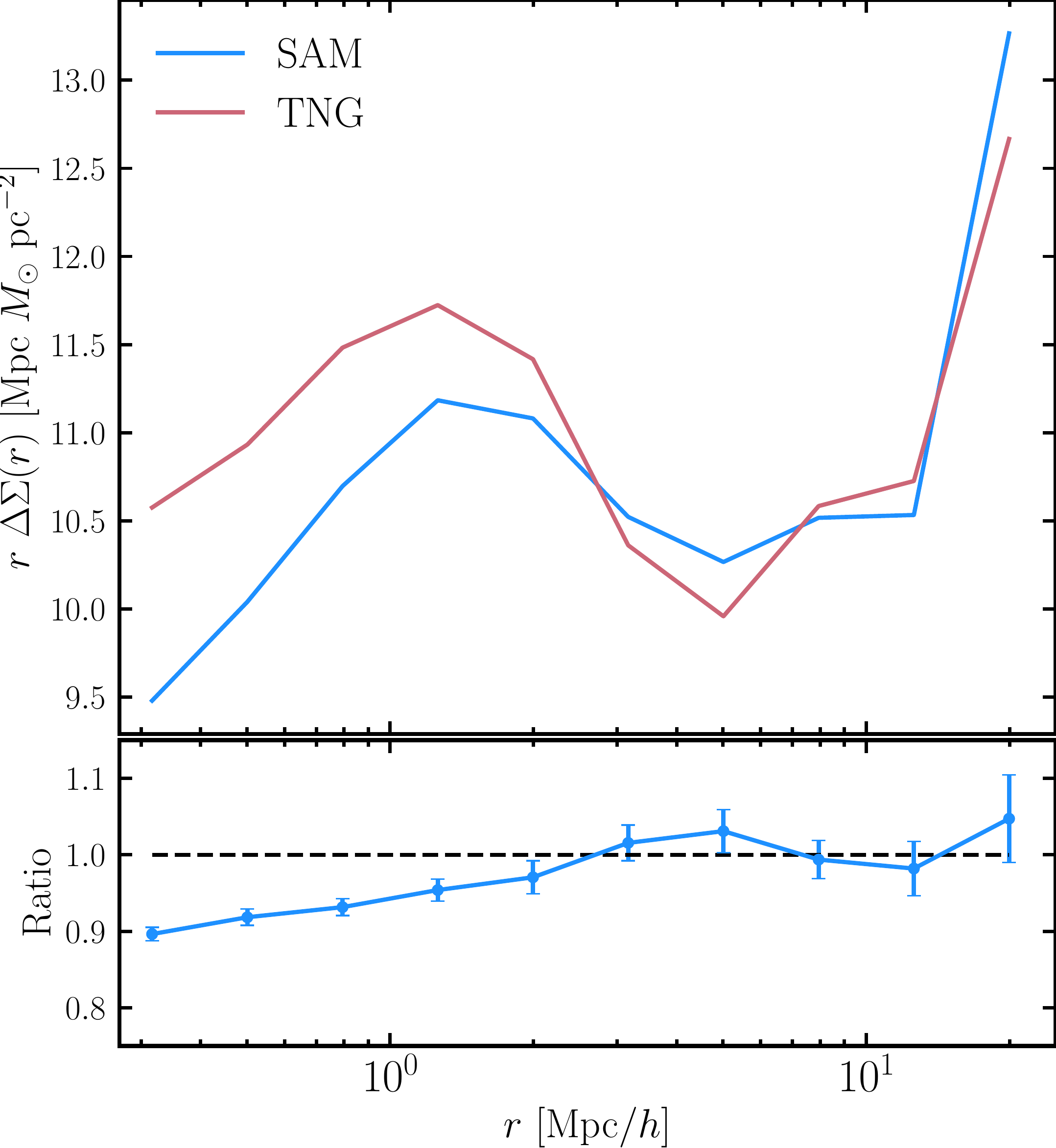}
\caption{Galaxy-galaxy lensing for a stellar-mass-selected sample with a galaxy number density of $n_{\rm gal} = 0.001 \ [{\rm Mpc}/h]^{-3}$ (corresponding to 12000 galaxies) at $z = 0$. In blue, we show the weak lensing signal of the SAM galaxy sample, while in red, we show the result for TNG. On large scales, the weak lensing signal appears to be consistent between the two models. On small scales, there is a discrepancy consistent with Fig.~\ref{fig:corrfunc}.}
\label{fig:gal_lens}
\end{figure}

As a measure of weak lensing, we consider the excess surface mass density profile, denoted as $\Delta \Sigma (r)$. It is obtained by first calculating
\begin{equation}
\Sigma(r_p) = \bar{\rho}\int_0^{\pi_{\rm max}} \left[1+\xi_{\rm gm}(\sqrt{r_p^2+\pi^2}) \right]\mathrm{d}\pi,
\end{equation}
where $\bar \rho$ is the mean matter density while $r_p$ and $\pi$ are the distances perpendicular and parallel to the line of sight, respectively. Then one can find the excess surface mass density as 
\begin{equation}
\Delta\Sigma(r_p) = \bar{\Sigma}(<r_p) - \Sigma(r_p),
\end{equation}
where the mean surface mass density interior to the projected radius is given by
\begin{equation}
\bar{\Sigma}(<r_p) = \frac{1}{\pi r_p^2}\int_0^{r_p}\Sigma(r_p^{\prime})2\pi r_p^{\prime} \mathrm{d}r_p^{\prime}.
\end{equation}

In Fig. \ref{fig:gal_lens}, we show the galaxy-galaxy lensing signal of the SC-SAM and TNG stellar-mass-selected samples at $z = 0$. On scales larger than $r \gtrsim 3 \ {\rm Mpc}/h$, the two are in good agreement with each other, whereas on smaller scales, they deviate by $\sim$10\%. This small-scale discrepancy should not come as a surprise given that their one-halo term contribution to the two-point clustering differs (see Fig.~\ref{fig:corrfunc}). The large scale consistency is encouraging, as it suggests that they have a similar connection, statistically speaking, to the underlying matter field. 
%\rss{the "lensing is low" comments got me wondering whether we should show observations on top of this -- i concluded that there probably are not observational results for the same redshift and mass selection that we are using -- is that correct? this could be an interesting follow-up study (to see if we can resolve the lensing/clustering tension with these more physical models) -- or has that been done?}

\subsubsection{Galaxy-void correlation function}

The study of cosmic voids offers an exquisite opportunity to study cosmology and understand what our Universe is made of. Voids are huge low-density regions ($10 - 100$ Mpc$/h$) that have undergone little non-linear growth  \citep{1978ApJ...222..784G}. Voids are complementary to both galaxy clustering and early-Universe measurements. They can help break existing degeneracies between cosmological parameters, as they are sensitive to a number of effects, such as redshift space distortions, baryon acoustic oscillations, neutrino effects, and the integrated Sachs-Wolfe effect \citep[e.g.]{2019MNRAS.488.4413K}. Cross-correlating the galaxy positions and the location of the voids offers insight into the relationship between large underdensities and densely clustered regions, exaggerating the most galaxy-deprived regions. It has the potential to differentiate between models that exhibit the same two-point auto-correlation statistics and thus help us generate more accurate mock catalogues to match the full statistical properties of the galaxy population.

\begin{figure}
\centering  
\includegraphics[width=0.48\textwidth]{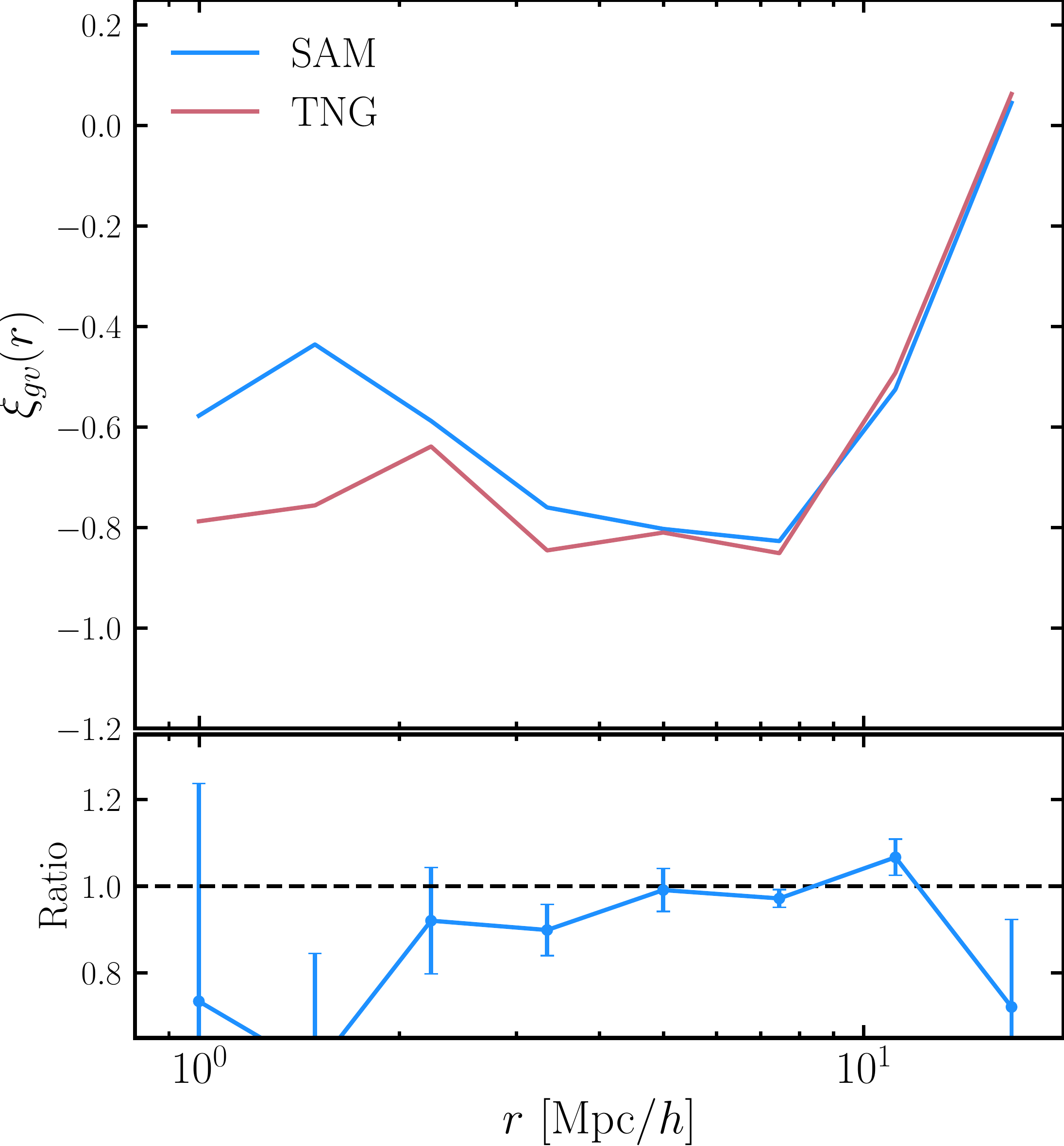}
\caption{Galaxy-void correlation function for a stellar-mass-selected sample with a galaxy number density of $n_{\rm gal} = 0.001 \ [{\rm Mpc}/h]^{-3}$ (corresponding to 12000 galaxies) at $z = 0$. In blue, we show the galaxy-void correlation function of the SAM galaxy sample, while in red, we show the result for TNG. On large scales, the two models show very good agreement, while on small scales, there is an evident discrepancy. Its significance is, however, difficult to assess due to the box limitations of the TNG simulation, so we do not place much importance on it.}
\label{fig:all_cross_voids}
\end{figure}

We use a simplified procedure for locating voids in the TNG simulation which was first outlined in \citet{2021MNRAS.501.1603H}:
\begin{enumerate}
\item Divide the TNG box into $128^3$ cubes of side $\sim 1.6$ Mpc$/h$.
\item Find the distance to the {\it third} nearest galaxy measured from the centre of each cube.
\item Order the obtained void candidates based on their size in descending order and for each object in the list, remove all voids whose centres lie within the boundaries of that object.
\item Record the void sizes and void centres for both the SAM and the TNG models.
\end{enumerate}
We identify on the order of 2000 voids, the smallest of which have radii $\sim$9 Mpc$/h$, while the largest reach 25 Mpc$/h$.

In Fig. \ref{fig:all_cross_voids}, we show the galaxy-void cross-correlation function for the TNG and SC-SAM stellar-mass-selected samples with $n_{\rm gal} = 0.001 \ [{\rm Mpc}/h]^{-3}$ at $z = 0$. On these scales, the cross-correlation between galaxies and voids is weak and negative, in agreement with previous works \citep{2021MNRAS.501.1603H}. On scales of $r \gtrsim 3 \ {\rm Mpc}/h$, the ratio of the two is consistent with one, but deviates on smaller scales. Given that the average size of the voids we have identified is $\sim$10 Mpc$/h$, we do not place much importance on this difference. As a result of the volume limitations of IllustrisTNG, we explore the cross-correlation function at significantly smaller scales than what is typically used in void studies ($r \sim 100 {\rm Mpc}/h$). This creates some difficulty in drawing robust conclusions from our analysis, and we defer more extensive comparisons at larger separation to future work.
%\rss{would it be interesting to also show the void probability function?}

\section{Discussion}
\label{sec:discussion}
Our results, along with other work in the literature, demonstrate that physical models of galaxy formation naturally predict a modulation of the galaxy HOD with parameters other than halo mass, which has a significant impact on clustering. However, this work taken together demonstrates that the degree of galaxy assembly bias, and its physical origin, is somewhat model dependent. One advantage of SAMs is that the models are relatively simple, so it is a bit easier to pull them apart and determine the physical origin of observed effects, unlike hydrodynamic simulations which are more difficult to interpret. For example, the SC-SAM does not currently include any \emph{explicit} environment-dependent physics in the models, so any observed dependence on environment must arise indirectly through the dependence of halo merger history on environment. In contrast, the galaxies in TNG experience different tidal forces in different environments, and may also be affected by ram pressure stripping, harrassment, or merging, all of which depend strongly on environment in a direct way. Indeed, we showed that the gas fraction in $10^{13} M_\odot$ halos in TNG shows significant dependence on environment. It is interesting therefore that for \emph{central} galaxies, the dependence of the HOD on environment in the SAMs is of a similar magnitude to that seen in TNG. However, the dependence of the \emph{satellite} HOD on environment is much stronger in TNG than in the SAMs, which is not surprising, as all of the environmental effects listed above are expected to work mainly on satellites. On the other hand, the SAMs show a much stronger dependence of the HOD on halo intrinsic properties, such as concentration and assembly time, than TNG. This may be because some of the SAM physics prescriptions explicitly contain these quantities or closely related ones, while this is not the case in TNG. For example, the mass outflow rate of stellar driven winds in TNG is an explicit function of the \emph{halo virial scale} velocity, while in the SAM, the corresponding quantity is parameterized in terms of the galaxy-scale \emph{disk velocity}, estimated as the halo rotation velocity at a radius of two NFW scale radii (which is directly related to concentration). Another example is that in the SAM, the galaxy disk scale radius is modeled using a formula that contains a direct dependence on both concentration and spin, while numerical simulations generally do not show a strong correlation between galaxy size and halo spin \citep[e.g.][]{2019MNRAS.488.4801J}. Work in progress (Karmakar et al. in prep) demonstrates that galaxy size and halo spin also show no significant correlation in the TNG simulations. 

The work presented here therefore points to the interesting result that galaxy-scale properties may be imprinted on large-scale galaxy clustering, through their impact on baryonic physics and the related influence on the HOD. We plan to use SAMs to further explore this by varying the parameterization of key physical processes such as stellar driven winds. Considered from another angle, it is clear that it is imperative to gain a better understanding of how to model the sub-grid physical processes in cosmological simulations, in a manner that faithfully captures the correct dependencies. Efforts such as SMAUG\footnote{Simulating Multiscale Astrophysics to Understand Galaxies; https://www.simonsfoundation.org/flatiron/center-for-computational-astrophysics/galaxy-formation/smaug/}, which aims to build up physical prescriptions for sub-grid models based on small-scale simulations that resolve the relevant physical processes, should help enable progress on this problem. 

An additional difference between the two models is that SC-SAM assumes that the hot gas halo is stripped instantaneously when a halo enters the virial radius and becomes a sub-halo. As a result, satellite galaxies have no source of gas accretion, and quickly consume their gas by turning it into stars or expel it via stellar driven winds. A more realistic picture is probably that the hot gas halos are gradually stripped over a finite timescale, while cold gas may also to some extent be stripped out via ram pressure and tidal forces. A further uncertainty regarding satellite evolution is that they may be tidally destroyed via interaction with (especially) the massive central galaxy as well as perhaps other satellites. In this study, we attempted to circumvent these issues by making the stellar-mass selections separately for centrals and satellites, and by placing the satellites in the SAM at the same locations as the satellites in TNG. However, in the future, SAMs should be updated to accurately account for these processes, using the results from numerical simulations as a guide. 

Another potential complication arises from the use of different halo catalogues -- the default halo finder for TNG is friends-of-friends (FoF) based, while the SAMs are built on halo merger trees based on the \textsc{ROCKSTAR} spherical overdensity based halo finder. We show that different halo finders can give rise to moderate differences in the assembly bias signatures for DM halos, complicating the interpretation of results in the literature which may have used different halo finders among other differences. However, we made use of bijective matches between the haloes generated by both methods, which allowed us to study the effect of this choice on our results. As shown in Fig.~\ref{fig:hod_rock}, this does not appear to affect our conclusions significantly.

In order to make an unbiased inference of cosmological parameters, it is essential that assembly bias is properly accounted for in the construction of mock catalogues. For example, \citet{2017MNRAS.467.3024L} showed that mock catalogs constructed with standard HOD techniques to match the clustering of the CMASS sample predict a galaxy-galaxy lensing signal that is 20-40\% larger than observed, and suggested that this could be due to shortcomings in our understanding of galaxy formation, or to new physics. Recent work has shown that modifying the standard HOD formalism to include assembly bias does reduce the ``lensing is low" discrepancy \citep[e.g.][]{2021MNRAS.502.3582Y,2021MNRAS.502.2074L,2020arXiv201001143Z}. The ability of relatively computationally efficient physics-based models, such as SAMs, to qualitatively reproduce the assembly bias signal seen in more sophisticated numerical hydrodynamic simulations is very promising for developing techniques to create large volume mock galaxy catalogs that are consistent with both lensing and galaxy clustering as well as other cosmological probes. 

In future work, we plan to retool the SC-SAMs to deliberately mimic the sub-grid physics recipes in the TNG simulations, including potentially tuning the calibration parameters to match TNG instead of observations. It is clear from the work presented here that attention should be paid to environment-dependent physical processes and especially their impact on satellite galaxies, such as tidal and ram pressure stripping, which will probably need to be included in the SAM. Other interesting questions to be explored include the dependence of assembly bias on redshift and on different tracers used to select galaxies, such as SFR or emission line based selection. Clearly our ultimate goal is to develop methods to produce mock catalogs with multiple observable tracers that reproduce the observed clustering properties over a broad range of redshift and for different tracers and galaxy selection criteria.

Finally, we recommend including more large-scale statistical probes in the validation and perhaps calibration of SAMs and simulations. We showed that although the SC-SAM and TNG produce very similar predictions for two-point statistics on large scales, they show interesting differences in higher order statistics such as the third moment of the smoothed density and void statistics. As we gain the ability to populate larger volumes with galaxies, observational measurements of these higher order clustering statistics will be invaluable for more stringent validation of the next generation of galaxy models. 

\subsection{Our results in the context of other studies in the literature} 

Some of the earliest studies on assembly bias in the literature have looked at the dependence of halo clustering on parameters beyond halo mass in dark-matter-only simulations \citep[e.g.][]{2005MNRAS.363L..66G,2007MNRAS.377L...5G}. \citet{2005MNRAS.363L..66G} found that haloes that assemble at early times are substantially more clustered than their late-time counterparts. In \citet{2007MNRAS.377L...5G}, the study was extended to include a larger number of parameters at different redshifts such as halo concentration and spin. These early papers were among the first to suggest that halo occupation depends on the large-scale environment of the halo and that assembly bias effects may lead to differences at the 10 per cent level.
%if one is to build galaxy population models that relate halo occupation solely to halo mass.
Our present study agrees qualitatively with the impact of assembly history on the halo clustering and confirms that in both TNG and SC-SAM, not accounting for secondary parameter dependencies leads to a discrepancy of order $10\%$.

More recent papers have explored the dependence of halo occupancy on additional halo parameters in different physics-based models of galaxy formation. For example, \citet{2018ApJ...853...84Z} investigate the effect of large-scale environment and halo formation time on galaxy samples obtained from a semi-analytic galaxy model. They find that early-forming haloes are more likely to host central galaxies at lower halo mass, while the opposite is true for the satellites. \citet{2018MNRAS.480.3978A}, on the other hand, perform a similar analysis, but this time employing the two hydrodynamical simulations, EAGLE and Illustris, reporting that halos in the densest environments are more likely to host a central galaxy than those in the least dense environments. This relationship is found to be even stronger for early-forming halos that have had more time to assemble. The present study similarly finds that halo occupation exhibits noticeable dependence on both environment and formation for TNG and SC-SAM, with TNG having more sensitivity to environment and SC-SAM on formation time.

Formation time and concentration, although extensively studied, have been shown, however, to be insufficient for reproducing the full galaxy assembly bias signal \citep[e.g.][]{2007MNRAS.374.1303C,2020MNRAS.493.5506H}. Several authors have also shown that augmenting galaxy population models with environmental information for each halo can help to fully reproduce the galaxy assembly bias signal \citep[e.g.][]{2021MNRAS.501.1603H,2021MNRAS.504.5205C,2021MNRAS.502.3242X}. \citet{sownak} offer a careful consideration of the secondary parameters affecting halo occupation in TNG and their complex interplay, whereas \citet{2018MNRAS.474.5143M} consider carefully the relationship between halo parameters in the context of the clustering of cluster-size haloes.

A work similar in spirit to ours was presented in \citet{2021MNRAS.504.5205C}, where the authors examined galaxy assembly bias on both a stellar-mass and a star-formation selected sample using TNG, a semi-analytic model, and a SHAM. While here we only focus on mass-selected populations, we put a significant effort into understanding the assembly bias properties of each model as well as the response of the galaxy clustering and halo occupation to a large variety of parameters. An additional difference is that our study includes statistics beyond the two-point correlation function such as moments of the density field and void statistics. An interesting observation is that SC-SAM and TNG appear to exhibit much more similar large-scale clustering behaviour and galaxy assembly bias than the SAM and hydro samples in \citet{2021MNRAS.504.5205C}. This may be attributed to differences in the SAMs employed in both studies and also the sample selection choices.

%\rss{papers to discuss: possibly `classic' halo assembly bias papers, such as Gao et al., Gao \& White, Mao et al.?; Zehavi et al. 2018 (HOD with classic SAMs), Contreras papers; Bose et al. 2019, Hadzhiyska et al. 2019, 2020; Artale et al. 2018 (occupancy variation for EAGLE and original Illustris).} 

\section{Summary and Conclusions}
\label{sec:conc}
In the last decade, cosmologists have been faced with the task of populating increasingly larger $N$-body simulations with galaxies to interpret the wealth of available and upcoming galaxy observations. There is established consensus that the most accurate way to do this involves \textit{ab-initio} prescriptions, which meticulously trace and evolve the various components of the Universe that contribute to galaxy formation according to the physical laws that govern them. Existing empirical recipes, such as the halo occupation distribution (HOD) and subhalo abundance matching (SHAM), make numerous simplifying assumptions; they not only lack physical motivation, but also are inaccurate at the 10\% level for certain tracers of the large-scale structure. On the other hand, semi-analytic models (SAMs) capture the essential physics of galaxy formation and output multiple galaxy properties, such as star formation rate and mass contained in stars, cold interstellar medium, and warm circumgalactic medium. They are thus a promising venue for modeling the connection between galaxies and the underlying matter distribution.

This work uses the latest version of the well-established Santa Cruz SAM, as run in merger trees extracted from the dark matter only version of the IllustrisTNG simulations \citep{Austen+2021}, and compares various large-scale properties of galaxies in the full-physics TNG simulations to those in the Santa-Cruz SAM (SC-SAM) at $z = 0$ for a stellar-mass-selected sample. We pointed out differences and similarities between the two models, and compared their assembly bias signatures.
%, and recommended ways to make the two more similar. 

The overall stellar-mass vs. halo mass (SMHM) relation and HOD for the SAMs and TNG are similar, but show some discrepancies. TNG tends to produce more massive galaxies in halos with $M_{\rm halo} \gtrsim 10^{13} \ M_\odot/h$ than the SC-SAM. This is due to differences in the AGN feedback physics recipes, and is partially a choice of calibration -- it is straightforward to reproduce more TNG-like results in the SC-SAM by reducing the strength of ``radio mode" AGN feedback. The HOD turns over more gradually in the SAM and has a longer ``tail" of galaxies populating lower mass halos. This is largely due to the sharper peak in the stellar-mass vs. halo mass relation and the larger scatter in this relation for halo masses around the peak ($M_{\rm halo} \gtrsim 10^{12} M_\odot$) in the SAM relative to TNG. Interestingly, SAMs seem to generically predict a larger scatter in the SMHM relation than numerical simulations \citep{2018ARA&A..56..435W}, for reasons that are not well-understood. The scatter in the SMHM relation at these halo masses is also not well constrained from observations, though results presented in \citet{2020MNRAS.498.5080C} suggest that the scatter increases towards lower halo masses in a manner more similar to the SAM predictions. 

A major finding is the excellent agreement between the SC-SAM and TNG for the two-point correlation function (Fig.~\ref{fig:corrfunc}), galaxy assembly bias signature (Fig.~\ref{fig:shuffle_all}) and qualitative response to the various halo parameters examined, seen in both the HOD as well as the galaxy assembly bias (Fig.~\ref{fig:hod_param_mstar} and Fig.~\ref{fig:shuffle_param_mstar}). This is despite the fact that the SAM was independently developed, with physical prescriptions that differ from those of TNG in many respects, and was calibrated to a different set of observations.  Such results are encouraging; they argue that the SC-SAM model to first order correctly captures many of the dominant features of the galaxy-halo connection produced by a much more computationally expensive method. Our results constitute direct evidence that SAMs are capable of capturing the dependence of clustering on halo parameters other than mass, thereby providing a physically motivated improvement over standard HOD models. 

However, there are some interesting and puzzling differences in the relationships between halo occupation and clustering with secondary parameters between the SAM and TNG. The SC-SAM and TNG central galaxy HODs show a similar degree of sensitivity to environment, but the SAMs show much greater sensitivity to other parameters such as concentration and halo formation time than does TNG. The clustering in both the SAM and TNG shows a significant sensitivity to environment when HODs are redistributed among halos that are ranked according to this parameter. TNG shows a weak sensitivity to internal parameters such as concentration and formation time in the ranked shuffling test, while clustering in the SAMs shows greater sensitivity to these parameters. As noted above, we speculate that this may be due to differences in the details of the quantities that are used to parameterize feedback recipes in the two models (e.g., halo scale vs. galaxy scale velocity).

In order to further explore the physical reasons for the differences seen between the models, we investigated the gas fractions on two different scales for halos of the same mass $M_{\rm halo} \sim 10^{13} \ M_\odot/h$ but in different environments in TNG, and compared this with the same quantity measured within the halo virial radius for SAMs. We found a significantly greater variation in gas fraction between low and high density environments in TNG than in the SAM, which showed virtually no dependence of this quantity on environment. 

Emulating state-of-the-art hydro simulations, such as IllustrisTNG, with ``cheaper'' methods that incorporate the basic galaxy formation model, such as SAMs, would revolutionise the development of  pipelines for cosmological surveys and change the face of precision cosmology. In the next few years, the extent to which SAMs can reproduce hydro simulations will be put to the test, and exciting advancements in cosmology are bound to come about!

\section*{Data Availability}
The IllustrisTNG data is publicly available at \url{www.tng-project.org}. 
The SAM outputs will be made available upon reasonable request to the authors. The scripts used in this project are available at \url{https://github.com/boryanah/SAM_TNG_clustering}.

\section*{Acknowledgements}
We would like to thank Sihan Yuan for illuminating discussions.
BH and DJE were supported by DOE-SC0013718 and NASA ROSES grant 12-EUCLID12-0004.
RSS gratefully acknowledges support from the Simons Foundation.
SB acknowledges support from Harvard University through the ITC Fellowship.
DJE is further supported in part as a Simons Foundation investigator.
Some of the simulations used in this work were produced using computing facilities provided by the Flatiron Institute. 

%%%%%%%%%%%%%%%%%%%%%%%%%%%%%%%%%%%%%%%%%%%%%%%%%%

%%%%%%%%%%%%%%%%%%%% REFERENCES %%%%%%%%%%%%%%%%%%

% The best way to enter references is to use BibTeX:

\bibliographystyle{mnras}
\bibliography{refs} % if your bibtex file is called example.bib

%%%%%%%%%%%%%%%%%%%%%%%%%%%%%%%%%%%%%%%%%%%%%%%%%%

%%%%%%%%%%%%%%%%% APPENDICES %%%%%%%%%%%%%%%%%%%%%

\appendix

%section{The halo finders of SC-SAM and TNG}
\section{Sensitivity to Halo Finders}
\label{app:halos}
In this section, we discuss the effects of utilizing two distinct halo-finding catalogues in our comparison between TNG and SC-SAM. As discussed in Section~\ref{sec:sam}, SC-SAM is constructed on top of a merger tree catalogue based on halos identified using the \textsc{ROCKSTAR} halo finder adopting the default \textsc{ROCKSTAR} parameters \footnote{For more information, see \url{https://github.com/yt-project/rockstar/blob/master/README.md}.}. TNG halo and subhalo properties, on the other hand, are reported in reference to catalogues obtained via the FoF and SUBFIND algorithms. It is known that different halo finders can yield different results for halo property distributions \citep{2011MNRAS.415.2293K,2011ApJ...740..102K}.

\begin{figure}
\centering  
\includegraphics[width=0.48\textwidth]{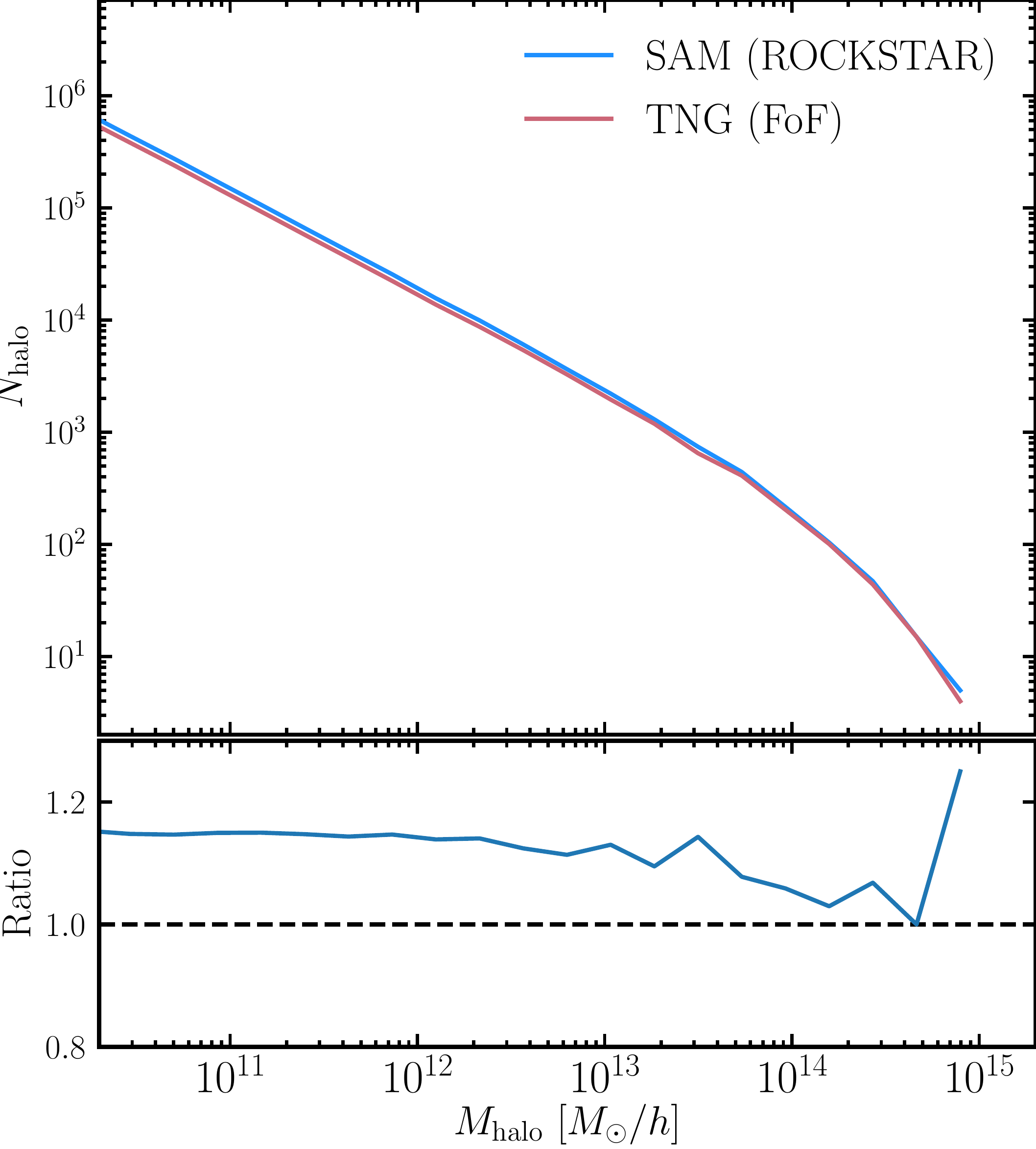}
\caption{Halo mass function (HMF) of halos identified with \textsc{ROCKSTAR} (blue) and with FoF/\textsc{SUBFIND} (red). We adopt the default virial mass definition from \textsc{ROCKSTAR}, while for FoF/subfind, we adopt `Group\_M\_TopHat200', which uses the same virial mass definition \citep{1998ApJ...495...80B}. The lower panel shows the ratio between the two and indicates that \textsc{ROCKSTAR} identifies 15\% more haloes on the mass scale between $10^{10}$ and ${10^13} \ M_\odot/h$.}
\label{fig:hmf}
\end{figure}

As an initial comparison, we show the halo mass function of the \textsc{ROCKSTAR} and FoF/\textsc{SUBFIND} catalogues in Fig. \ref{fig:hmf}. As halo mass proxy in TNG, we adopt `Group\_M\_TopHat200', as it uses the same virial mass definition \citep{1998ApJ...495...80B} as the default choice in \textsc{ROCKSTAR}. Despite that, we see that the discrepancy in the number of haloes between the two catalogues is $\sim$15\% for halo masses between $10^{10}$ and $10^{13} \ M_\odot/h$. At higher halo masses, the difference is smaller, and the ratio approaches unity for the largest haloes. The most likely explanation for the number of \textsc{ROCKSTAR} haloes being larger is that the \textsc{ROCKSTAR} finder uses 6-dimensional information to identify haloes and thus deblends many of the large percolating FoF groups. This is particularly noticeable at higher densities, where the FoF finder tends to string together distinct dark-matter structures into a single FoF object particularly when they are orbiting on the outskirts of large dark-matter clumps.

In an effort to correct for the fact that TNG reports FoF halo properties, in Fig.~\ref{fig:hod_param_mstar} and Fig.~\ref{fig:shuffle_param_mstar}, we make use of the bijective matches provided between the SC-SAM and TNG catalogues. In addition, here we explore the effect of associating the TNG galaxies with \textsc{ROCKSTAR} haloes. We do this via the following procedure: for each TNG subhalo, we find the most massive \textsc{ROCKSTAR} halo that encompasses it within its virial radius and assign it to that halo. For the stellar-mass-selected sample at $z = 0$, we find matches for 99.7\% of the TNG galaxies. We see that that the HOD of the TNG galaxies remains almost unchanged. This implies that the root cause of the differences between TNG and the SAM is likely unrelated to the halo finding, but is rather caused by a different implementation of the galaxy formation processes and the resulting galaxy-halo connection.

\begin{figure}
\centering  
\includegraphics[width=0.48\textwidth]{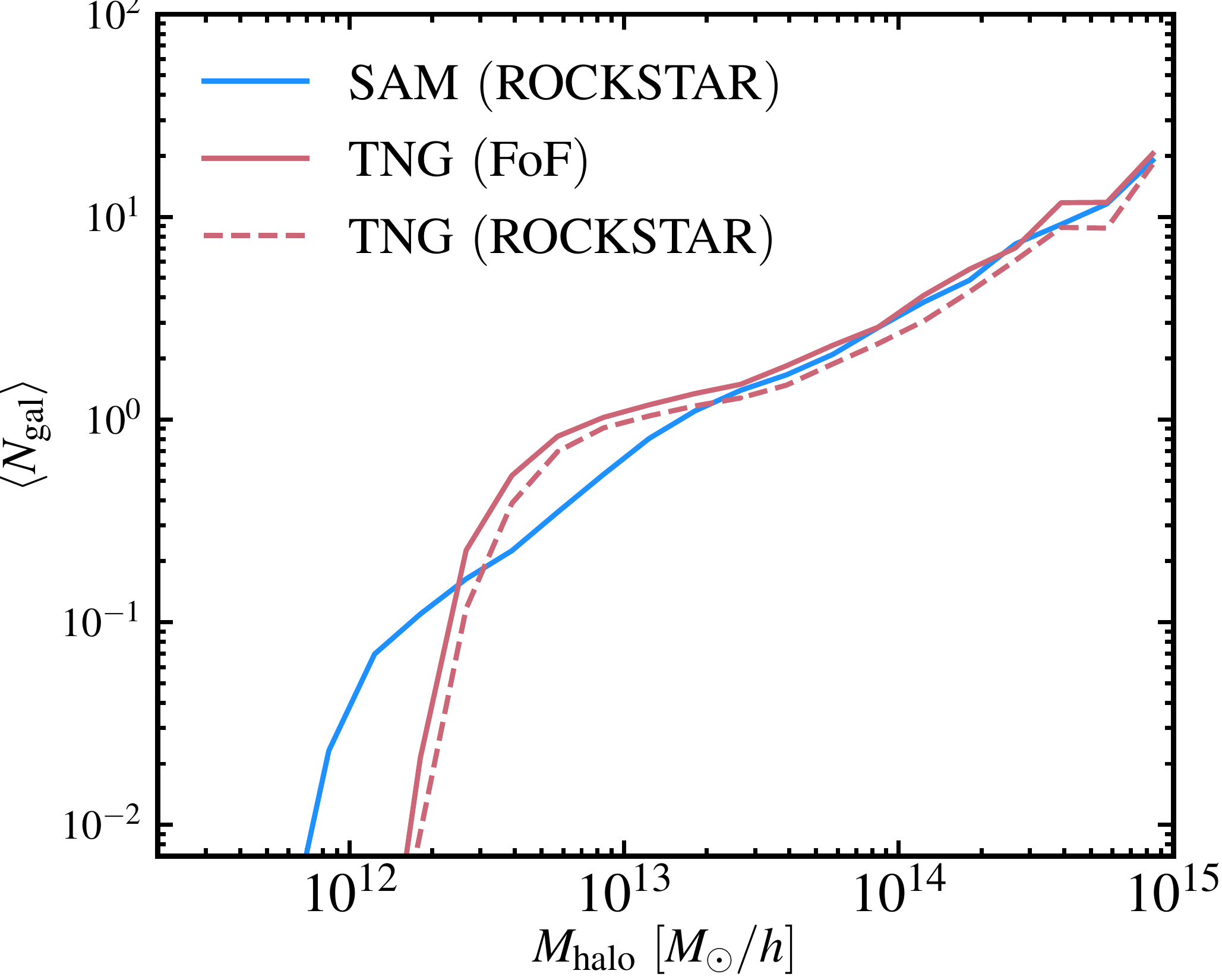}
\caption{Halo occupation distribution for a stellar-mass-selected sample with a galaxy number density of $n_{\rm gal} = 0.001 \ [{\rm Mpc}/h]^{-3}$ (corresponding to 12000 galaxies) at $z = 0$ at $n_{\rm gal} = 0.001 \ [{\rm Mpc}/h]^{-3}$. In blue, we show the HOD of the SAM galaxy sample, which is constructed from ROCKSTAR halo catalogues, while in red, we show the result for TNG. The default halo finder used to compute the halo statistics in TNG is FoF/\textsc{SUBFIND}. In dashed red, we show what the HOD would look like if we match the TNG galaxies to ROCKSTAR haloes and use the ROCKSTAR based mass instead. We see that the HOD of the TNG galaxies changes only marginally.}
\label{fig:hod_rock}
\end{figure}

\section{Halo assembly bias and parameter correlations}
\label{app:hab_corr}
In this section, we study the strength of the correlation between various halo parameters that are relevant in the study of galaxy assembly bias. We also visualize the halo assembly bias signal resulting from halo concentration and environment. These relations have been extensively studied and reported on in the literature, so we present them here largely as a baseline. 

\begin{figure*}
\centering  
\includegraphics[width=0.39\textwidth]{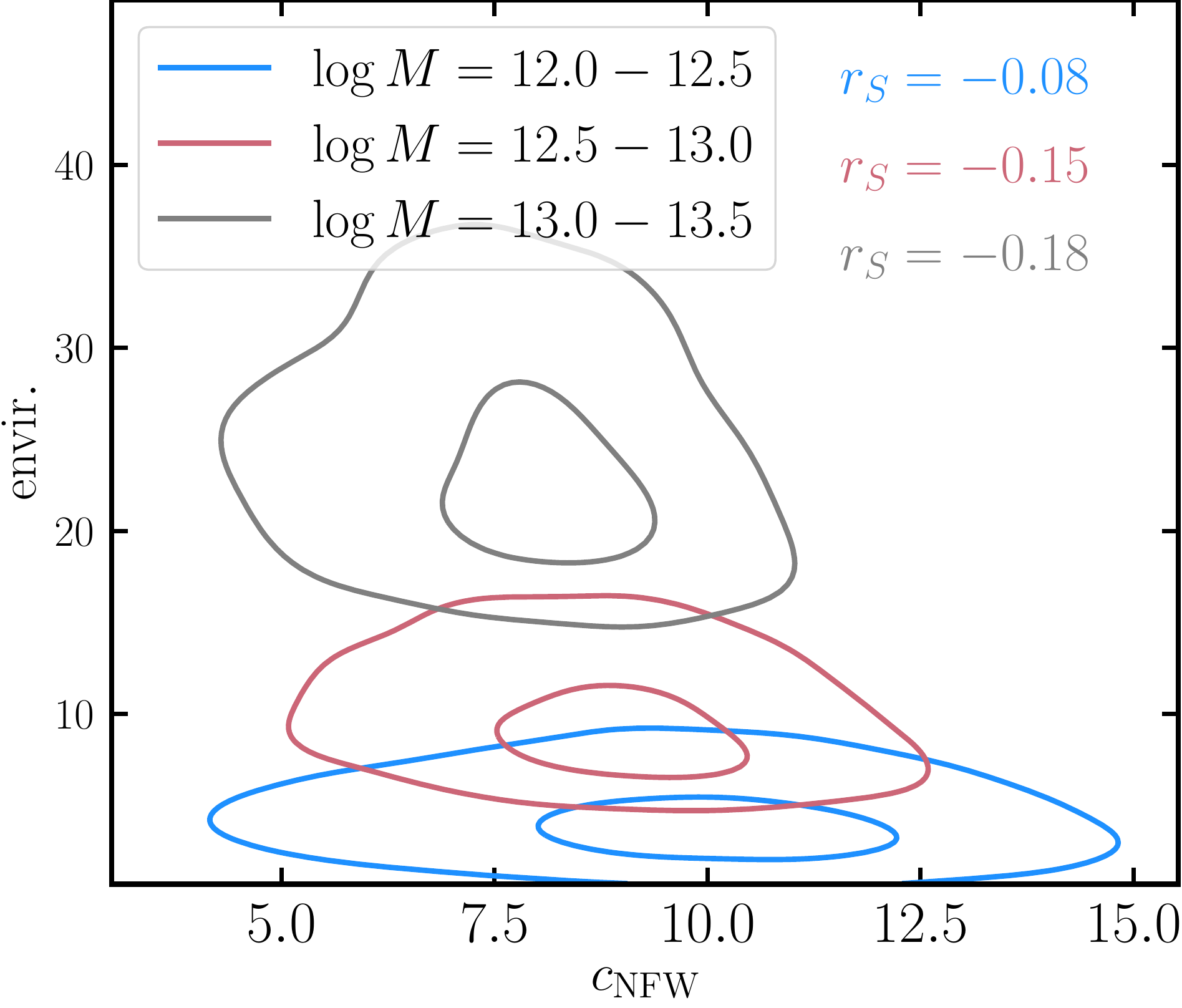}\hspace{0.1\textwidth}
\includegraphics[width=0.39\textwidth]{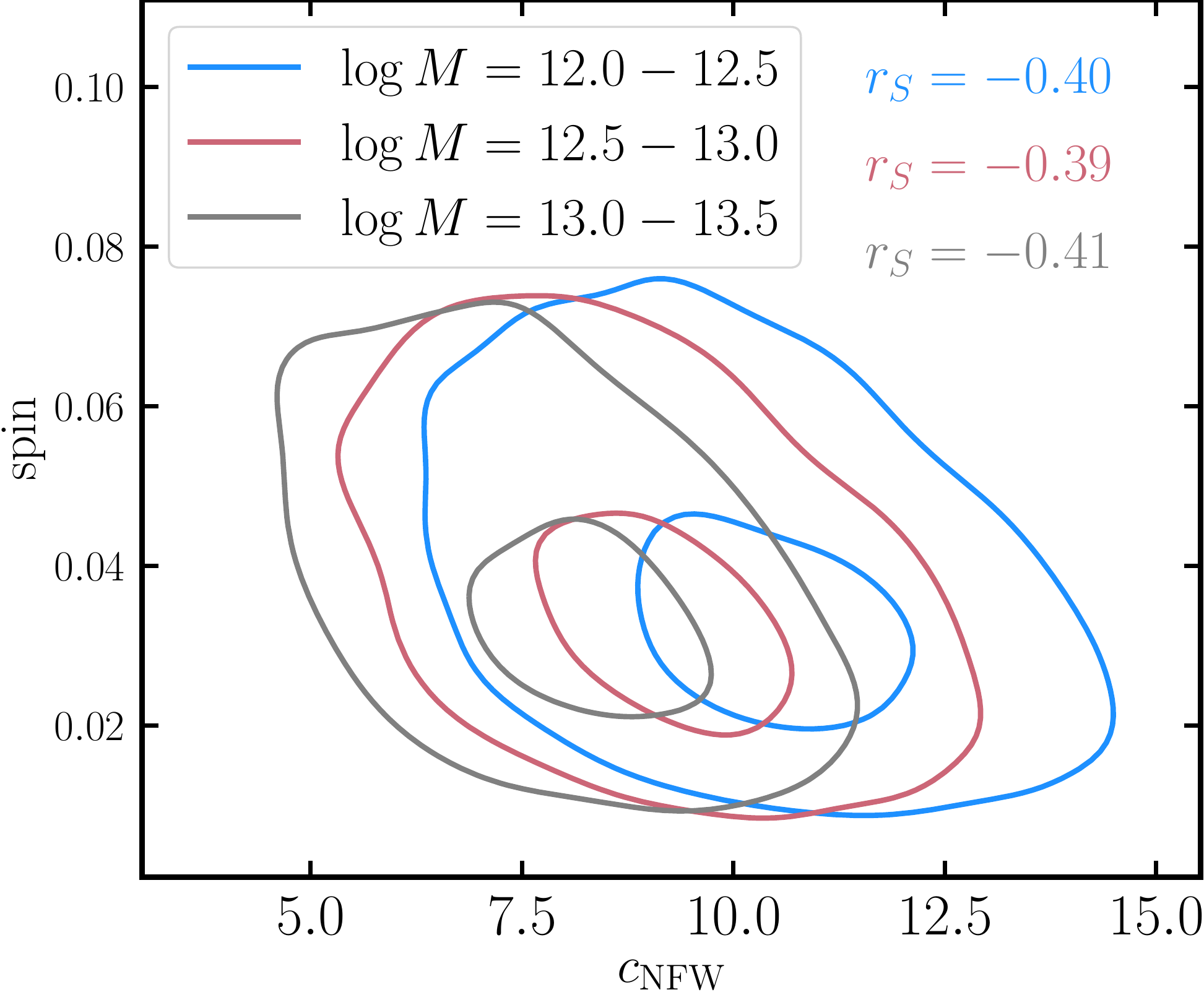}
\includegraphics[width=0.39\textwidth]{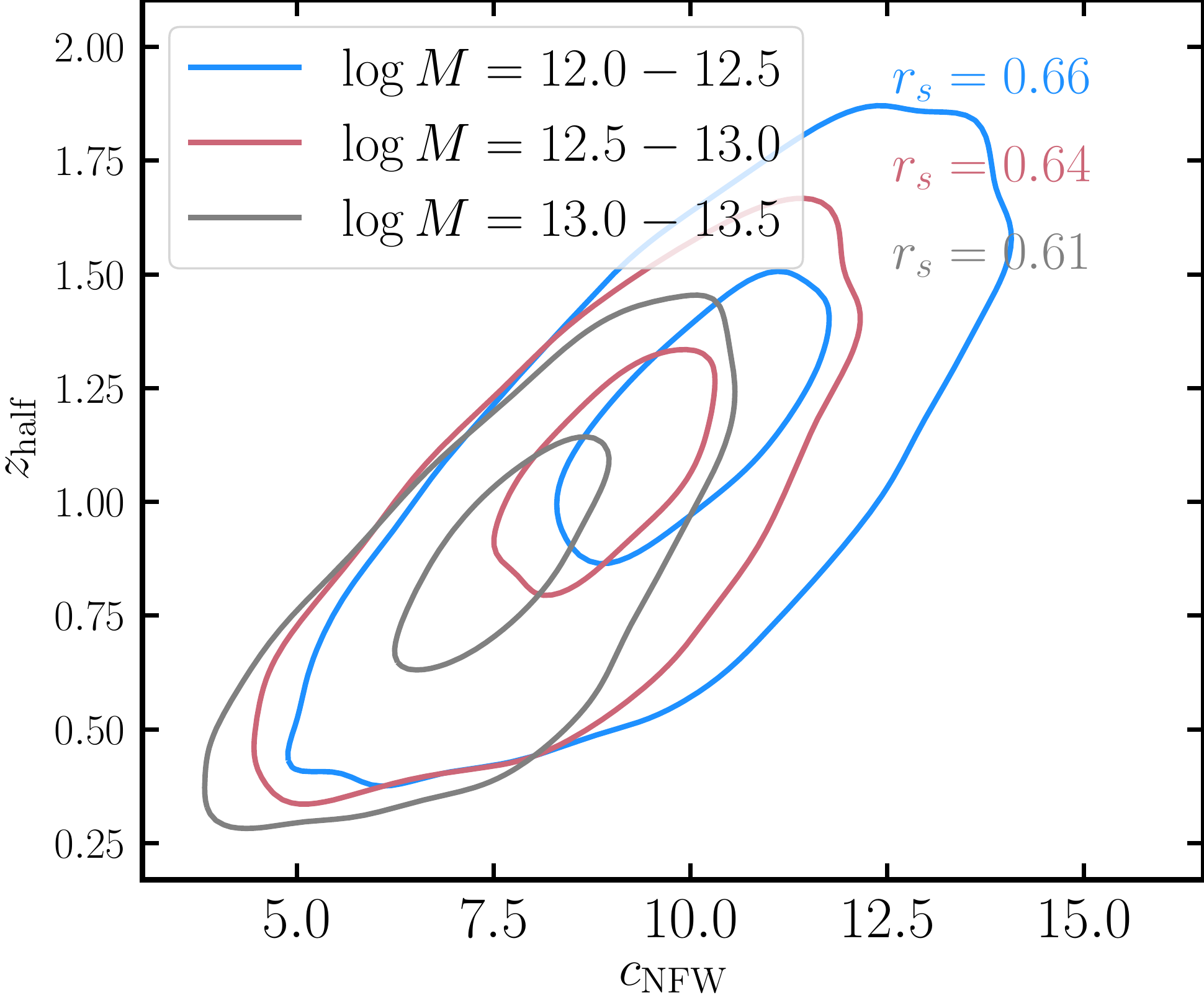}\hspace{0.1\textwidth}
\includegraphics[width=0.39\textwidth]{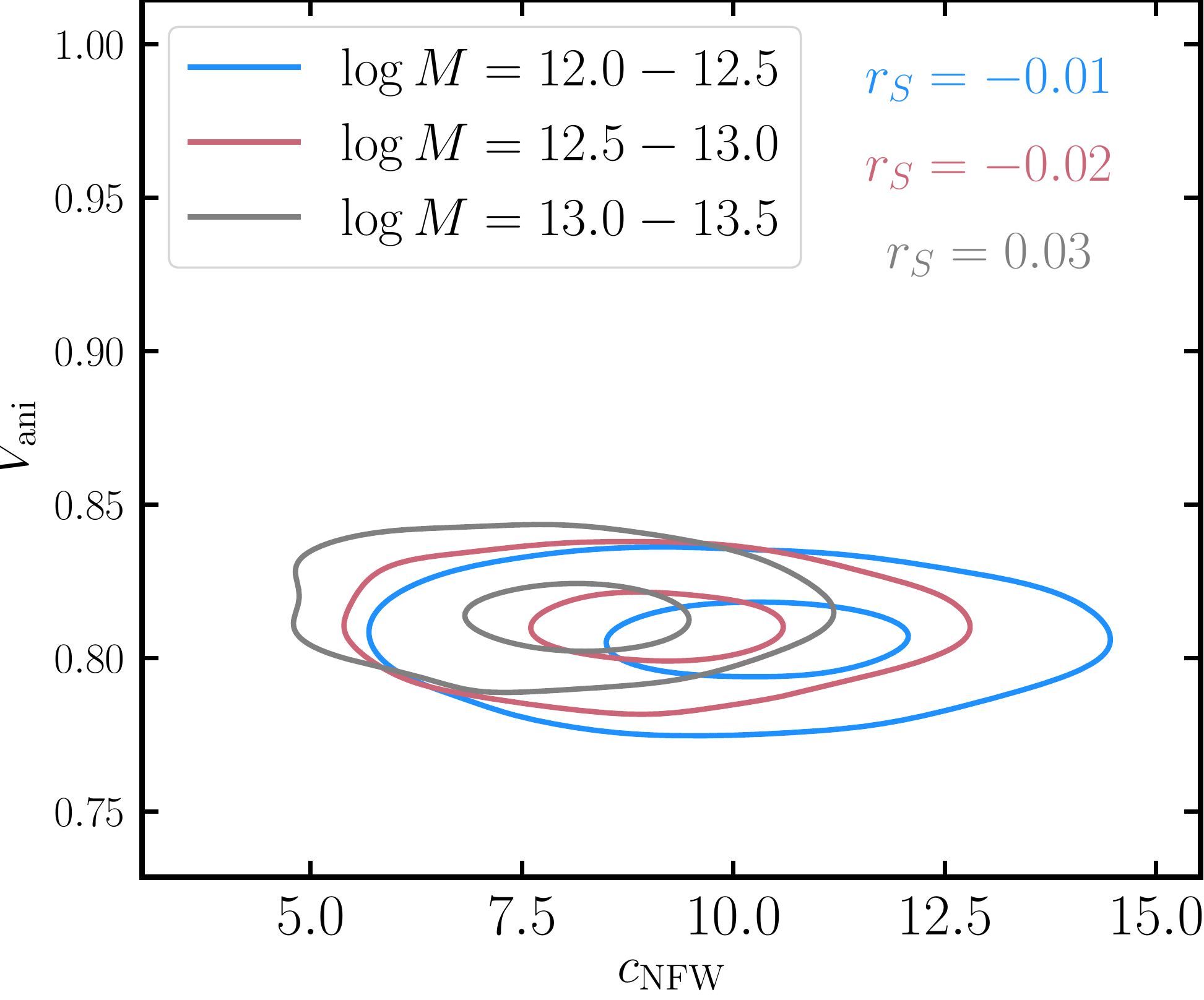}
\includegraphics[width=0.39\textwidth]{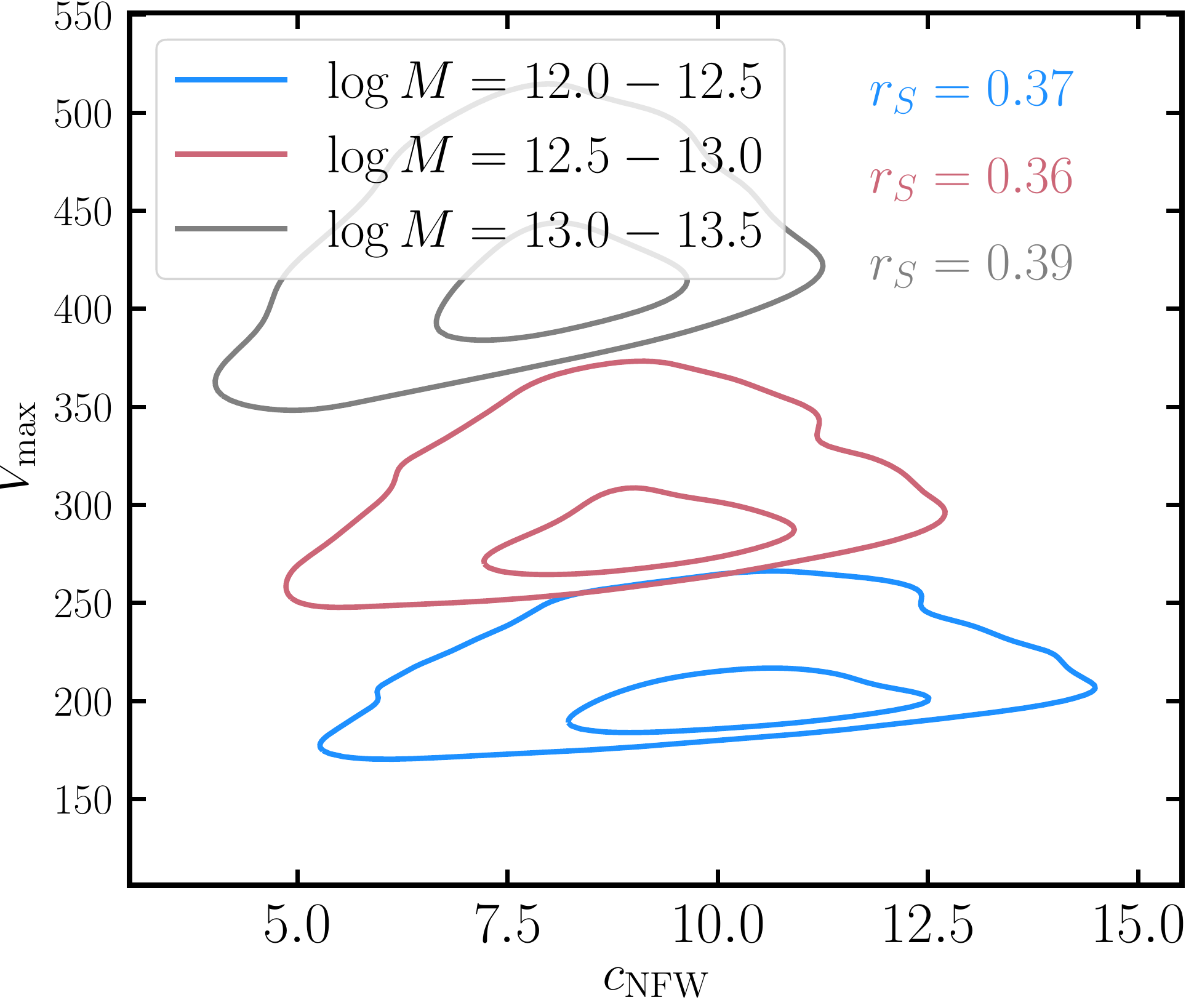}\hspace{0.1\textwidth}
\includegraphics[width=0.39\textwidth]{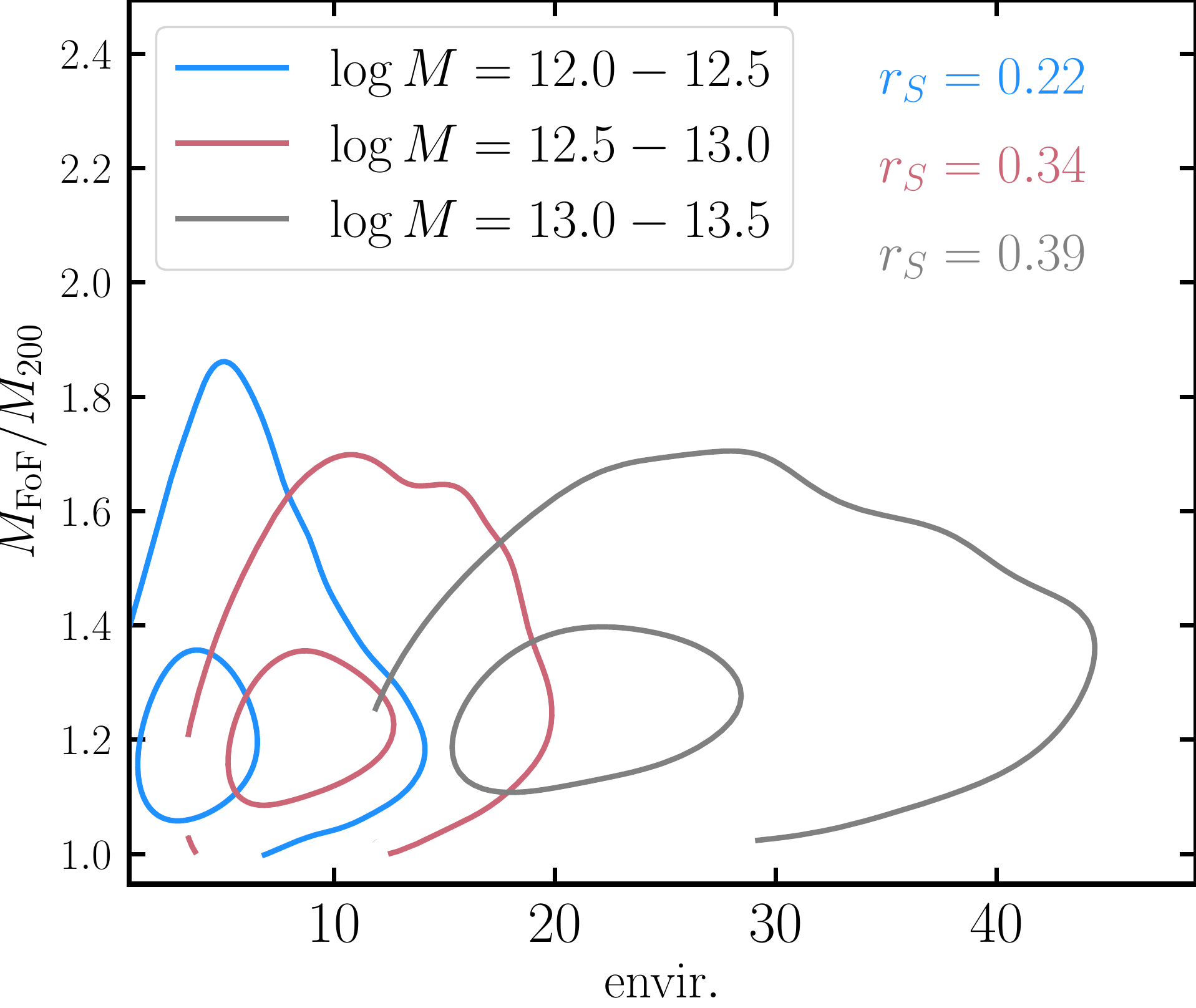}
\includegraphics[width=0.39\textwidth]{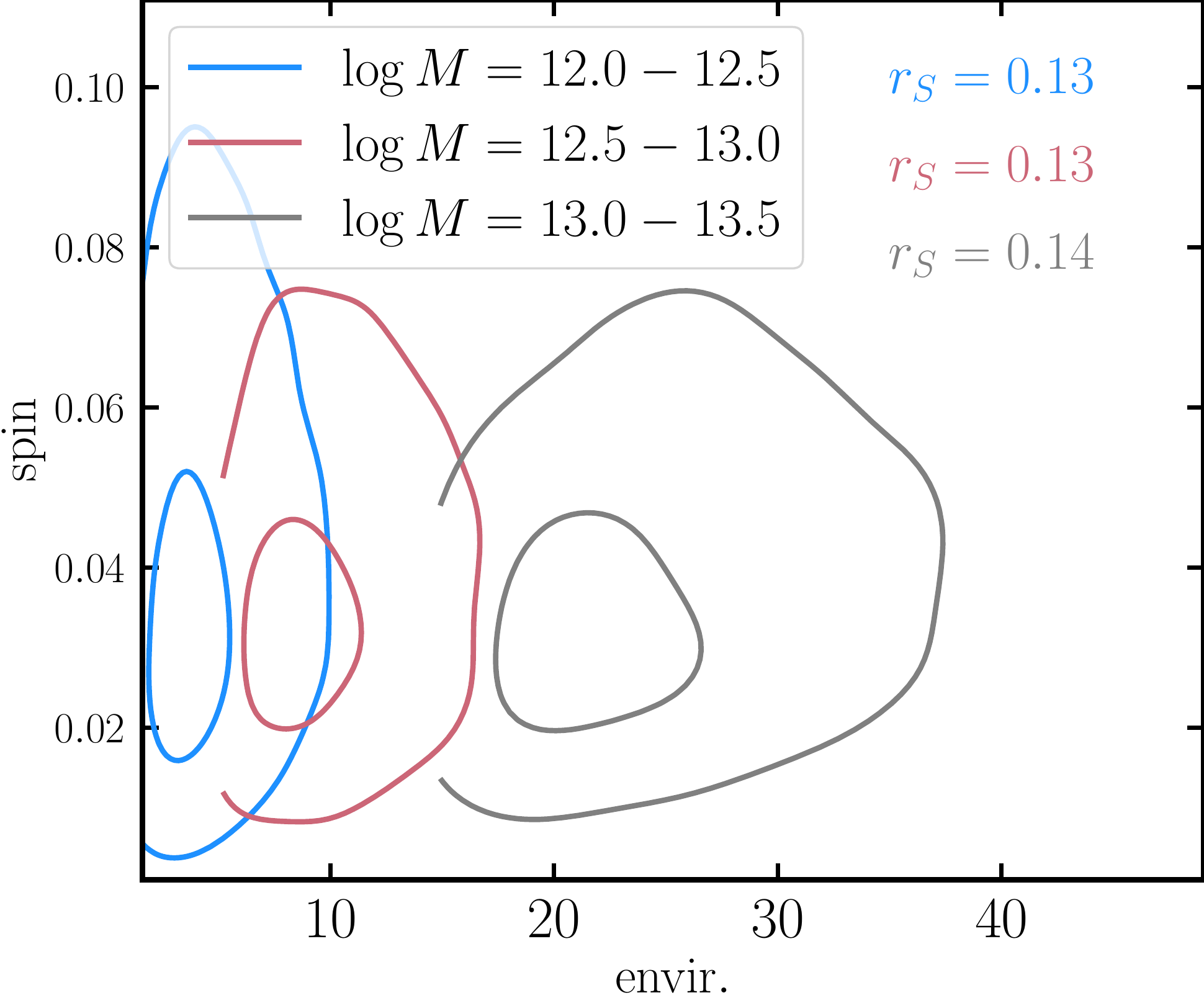}\hspace{0.1\textwidth}
\includegraphics[width=0.39\textwidth]{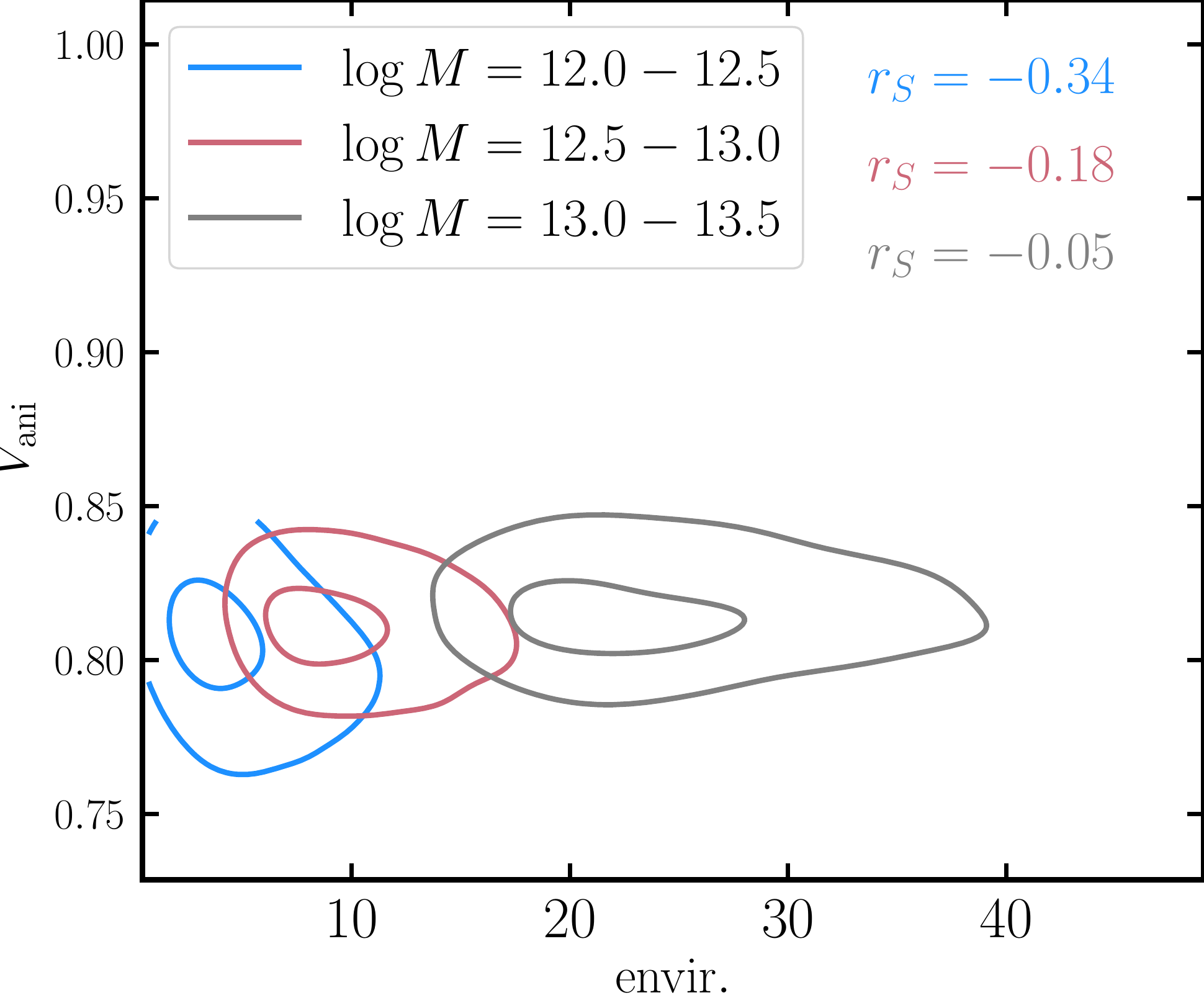}
\caption{Correlations between the different halo parameters for 3 mass bins ($\log M_{\rm halo} = 12.0 - 12.5, \ 12.5 - 13.0, \ {\rm and} \ 13.0 - 13.5$) in units of $M_\odot/h$. Spearman's rank coefficients, $r_S$, are shown for each bin in colors corresponding to their respective mass bins. Higher $r_S$ values correspond to stronger correlations. The density contours correspond to the 70th and 95th percentile of the KDE-weighted data, respectively.
%\rss{it would be good to adjust the y-axes to better 'hug' the data plotted. also, the labels in the legend on the top plot do not make sense. add more description of what is plotted to the caption so that figure can be read without referring to the text. }
}
\label{fig:halo_corr}
\end{figure*}

In Fig.~\ref{fig:halo_corr},
%\rss{is this figure created using the FoF/subfind based halos and parameters? does it look any different if you use ROCKSTAR? for both B1 and B2, are you plotting only "distinct" halos or subhalos as well (though there won't be many of them at these masses)?}
we show 70\% and 95\% two-dimensional contours, illustrating the scatter plots of a few pairs of halo parameters, weighted using kernel density estimation (KDE). More information about the parameters, including their definitions (environment, concentration, formation epoch, spin, velocity anisotropy, maximum circular velocity, FoF to virial mass ratio), can be found in Section~\ref{sec:params}. The haloes are split into three mass ranges: $\log M_{\rm halo} = 12.0 - 12.5, \ 12.5 - 13.0, \ {\rm and} \ 13.0 - 13.5$, in units of $M_\odot/h$. We also quote the Spearman's rank coefficient, $r_S$, for each mass bin \citep{spearman04}. 
%\rss{slightly unfortunate notation, since this also refers to the NFW scale radius...}
To get a sense of the correlation strength, we remind the reader that the ranges $0 \leq r_S < 0.2$, $0.2 \leq r_S < 0.4$, $0.4 \leq r_S < 0.6$, and $0.6 \leq r_S < 0.8$ correspond to very weak, weak, moderate, and strong correlation. The strongest correlation is between the formation time and concentration, where as expected early-forming haloes are much more concentrated. The correlation between spin and concentration is also relatively stronger compared with that between other pairs and confirms the findings of other authors \citep[e.g.][]{sownak}. Interestingly, environment is very weakly correlated with concentration and spin, so including it as an assembly bias parameter would not introduce a strong degeneracy in the halo model. Environment demonstrates the strongest amount of correlation with velocity anisotropy for low-mass haloes and the ratio of the FoF to virial mass for high-mass haloes. A likely reason for this finding is that high-environment haloes live in dense regions with a large number of mergers (hence lower velocity anisotropy), where the FoF group percolates over a large distance.

\begin{figure*}
\centering  
\includegraphics[width=0.93\textwidth]{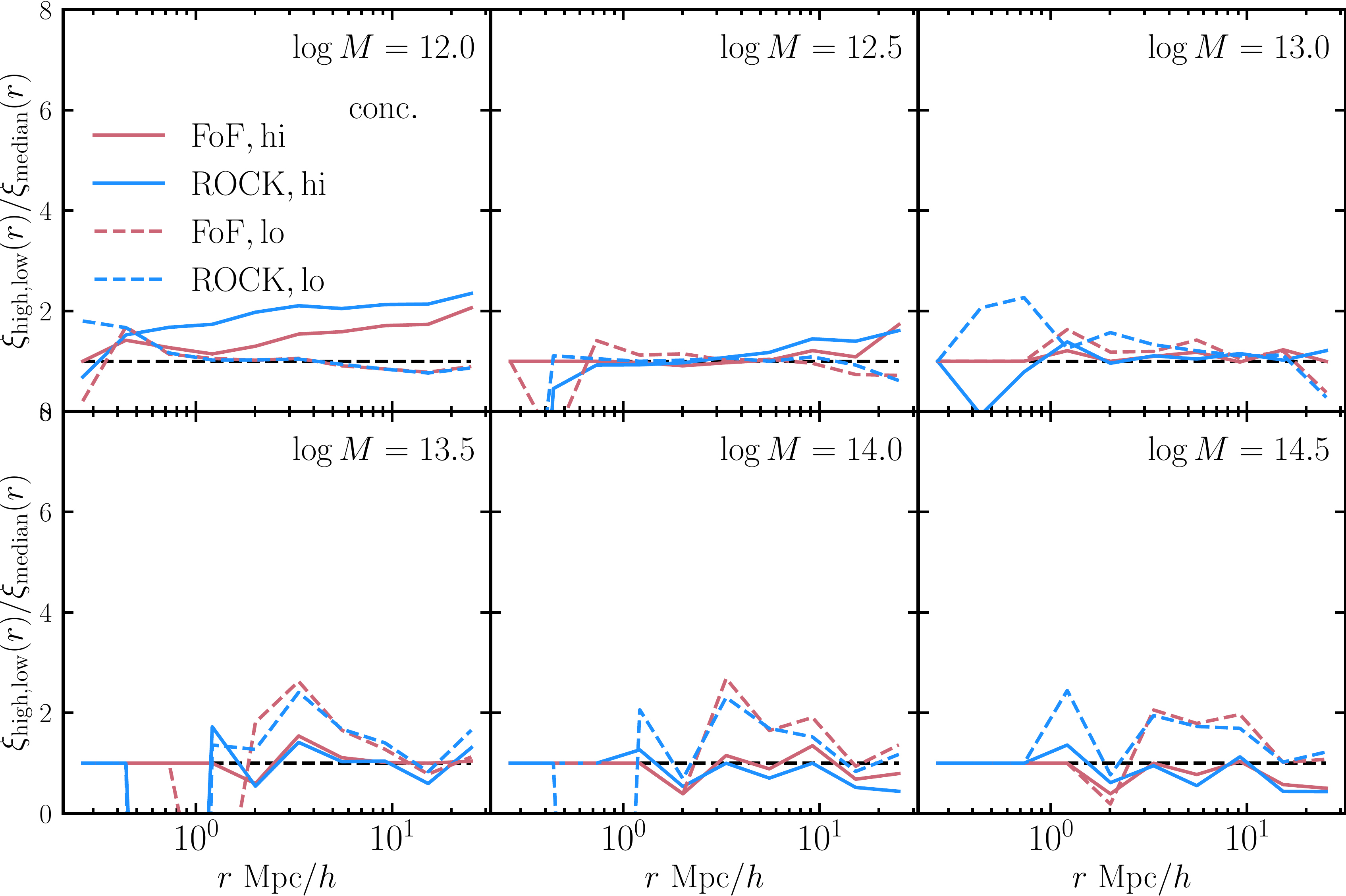} \\
\includegraphics[width=0.93\textwidth]{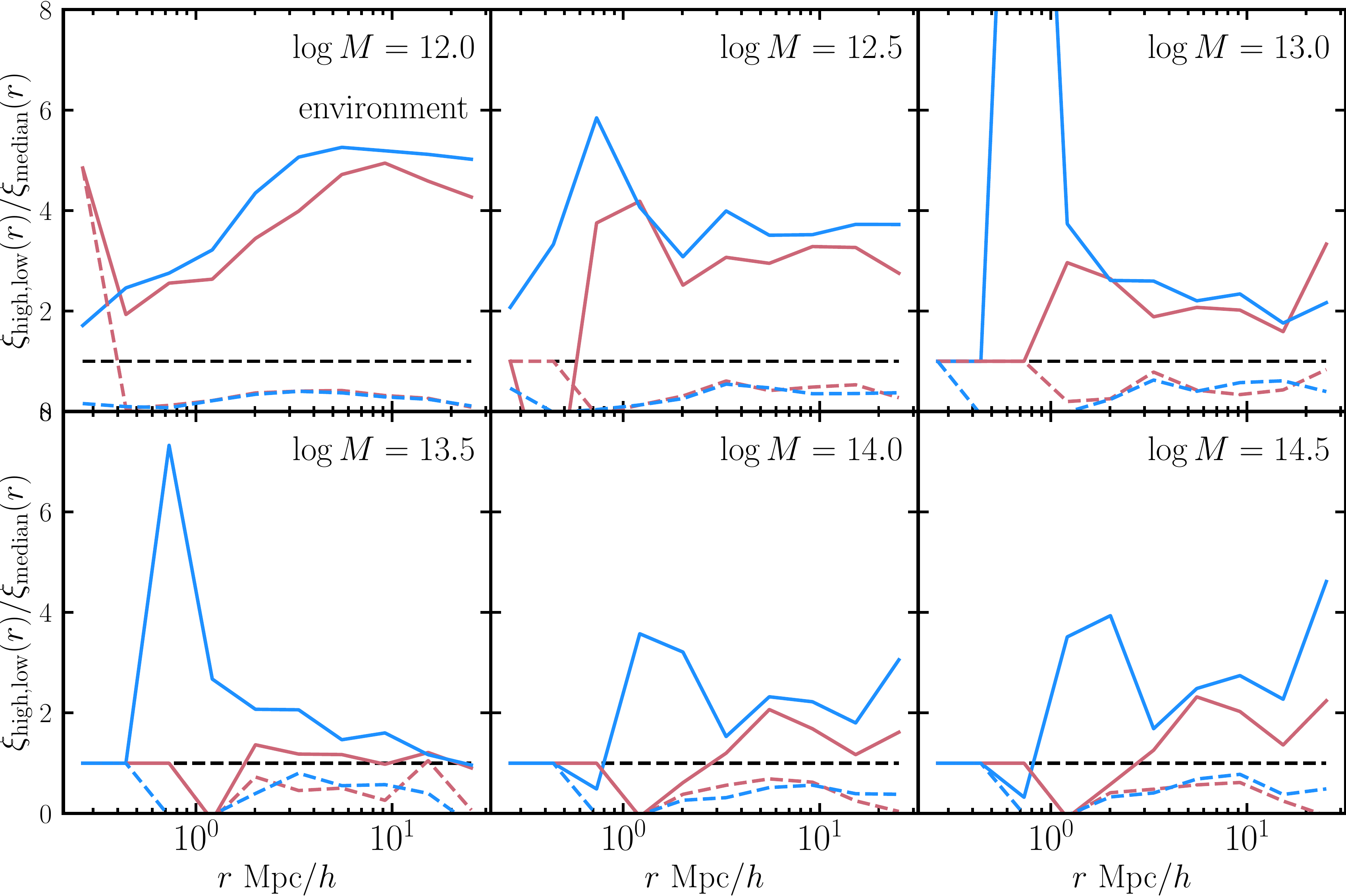}
\caption{
%\rss{this is just for halos -- so there is no SAM or TNG! just use the labels rockstar and FoF}.
Halo assembly bias signature for the concentration (top) and environment (bottom) parameters for different halo masses ($\log M = 12.0, \ 12.5, \ 13.0, \ 13.5, \ 14.0, \ 14.5$).  In blue, we show the halo assembly bias of the ROCKSTAR haloes ('ROCK'), while in red, we show the result for FoF. The dotted lines correspond to the bottom 30\% of haloes ('lo'), whereas the solid lines denote the top 30\% ('hi'), when ordered by concentration/environment. The ROCKSTAR haloes are abundance matched to the TNG-selected haloes for each mass bin to make the comparison more direct. All masses are reported in units of $M_\odot/h$. The signal is much weaker in the concentration plot and for small halo masses, and we see that high concentration leads to positive assembly bias, whereas on the high mass end, the relation is inverted. On the other hand, in the case of environment, the bias signal is always stronger for haloes in high-density environments and weaker for haloes residing in low-density regions.}
\label{fig:hab}
\end{figure*}

To estimate the halo assembly bias, we split the haloes into 6 mass bins: $\log M = 12.0\pm0.3, \ 12.5\pm0.3, \ 13.0\pm0.3, \ 13.5\pm0.3, \ 14.0\pm0.4, \ 14.5\pm0.5$. We define halo assembly bias as the ratio between the correlation function of the top/bottom 25\% haloes ordered by the parameter of choice, divided by the correlation function of the median population ($50\%\pm12.5\%$). The ROCKSTAR haloes are abundance matched to the TNG-selected haloes for each mass bin to make the comparison more direct. 
%\rss{i am not sure what it means to say that they are "abundance matched" given that you are using mass bins}
The resulting comparison is shown in Fig.~\ref{fig:hab}. For small haloes with $M_{\rm halo} \sim 10^{12} \ {M_\odot/h}$, we see that high concentration results in stronger clustering, whereas this relation is reversed for the most massive haloes as expected based on previous studies in the literature. On the other hand, in the case of environment, across all mass bins haloes living in denser environments are more clustered than haloes living in less dense regions. As discussed in the main text, haloes in dense regions experience more mergers and can be part of extended cosmic web structures such as walls, filaments, and knots that have typical size of $\sim$10 Mpc$/h$. Comparing TNG with SAM, we see that the signal is stronger for SAM, confirming the conjecture that \textsc{ROCKSTAR} tends to deblend haloes more readily and identify dark matter structure on the outskirts of large haloes. It is also noteworthy that the halo assembly bias response to environment is significantly stronger than the response to concentration.

\iffalse
\section{Stellar-to-halo mass ratio}
\begin{figure*}
\centering  
\includegraphics[width=1\textwidth]{shmr_param_mstar}
\caption{Dependence on the halo parameters of the stellar-to-halo mass ratio for a stellar-mass-selected sample with a galaxy number density of $n_{\rm gal} = 0.001 \ [{\rm Mpc}/h]^{-3}$ (corresponding to 12000 galaxies) at $z = 0$. In solid blue, we show the SHMR of the SAM galaxy sample, while in solid red, we show the result for TNG.}
\label{fig:shmr_param_mstar}
\end{figure*}
\begin{figure*}
\centering  
\includegraphics[width=1\textwidth]{shmr_all}
\caption{Stellar-to-halo mass ratio for a stellar-mass-selected (top) and a SFR-selected (bottom) galaxy sample with a galaxy number density of $n_{\rm gal} = 0.001 \ [{\rm Mpc}/h]^{-3}$ (corresponding to 12000 galaxies) at $z = 0$. In solid blue, we show the SHMR of the SAM galaxy sample, while in solid red, we show the result for TNG.}
\label{fig:shmr_all}
\end{figure*}
\begin{figure*}
\centering  
\includegraphics[width=1\textwidth]{shmr_all_55}
\caption{Stellar-to-halo mass ratio for a stellar-mass-selected (top) and a SFR-selected (bottom) galaxy sample with a galaxy number density of $n_{\rm gal} = 0.001 \ [{\rm Mpc}/h]^{-3}$ (corresponding to 12000 galaxies) at $z = 0.8$. In solid blue, we show the SHMR of the SAM galaxy sample, while in solid red, we show the result for TNG.}
\label{fig:shmr_all_55}
\end{figure*}
\fi

%%%%%%%%%%%%%%%%%%%%%%%%%%%%%%%%%%%%%%%%%%%%%%%%%%

% Don't change these lines
\bsp	% typesetting comment
\label{lastpage}
\end{document}